THESIS FOR THE DEGREE OF DOCTOR OF PHILOSOPHY

# Construction of force measuring optical tweezers instrumentation and investigations of biophysical properties of bacterial adhesion organelles

MAGNUS ANDERSSON

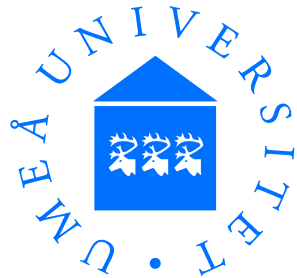

DEPARTMENT OF PHYSICS

UMEÅ UNIVERSITY

2007

Construction of force measuring optical tweezers instrumentation and investigations of biophysical properties of bacterial adhesion organelles

Department of Physics
Umeå University
SE-901 87 Umeå
Sweden

To my daughter Nea

## ABSTRACT

Optical tweezers are a technique in which microscopic-sized particles, including living cells and bacteria, can be non-intrusively trapped with high accuracy solely using focused light. The technique has therefore become a powerful tool in the field of biophysics. Optical tweezers thereby provide outstanding manipulation possibilities of cells as well as semi-transparent materials, both non-invasively and non-destructively, in biological systems. In addition, optical tweezers can measure minute forces ($< 10^{-12}$ N), probe molecular interactions and their energy landscapes, and apply both static and dynamic forces in biological systems in a controlled manner. The assessment of intermolecular forces with force measuring optical tweezers, and thereby the biomechanical structure of biological objects, has therefore considerably facilitated our understanding of interactions and structures of biological systems.

Adhesive bacterial organelles, so called pili, mediate adhesion to host cells and are therefore crucial for the initial bacterial-cell contact. Thus, they serve as an important virulence factor. The investigation of pili, both their biogenesis and their expected *in vivo* properties, brings information that can be of importance for the design of new drugs to prevent bacterial infections, which is crucial in the era of increased bacterial resistance towards antibiotics.

In this thesis, an experimental setup of a force measuring optical tweezers system and the results of a number of biomechanical investigations of adhesive bacterial organelles are presented. Force measuring optical tweezers have been used to characterize three different types of adhesive organelles under various conditions, P, type 1, and S pili, which all are expressed by uropathogenic *Escherichia coli*. A quantitative biophysical force-extension model, built upon the structure and force response, has been developed. It is found, that this model describes the biomechanical properties for all three pili in an excellent way. Various parameters in their energy landscape, e.g., bond lengths and transition barrier heights, are assessed and the difference in behavior is compared. The work has resulted in a method that in a swift way allows us to probe different types of pili with high force and high spatial resolution, which has provided an enhanced understanding of the biomechanical function of these pili.

Sammanfattning

Optisk pincett är en teknik i vilken mikrometerstora objekt, inkluderande levande celler och bakterier, beröringsfritt kan fångas och förflyttas med hög noggrannhet enbart med hjälp av ljus. Den optiska pincetten har därmed blivit ett kraftfullt verktyg inom biofysiken, som möjliggör enastående precisions-manipulering av celler och semi-transparenta objekt. Dessutom kan denna manipulation göras intracellulärt, dvs. utan att fysiskt öppna eller penetrera cellernas membran. Den optiska pincetten kan även mäta mycket små krafter och interaktioner ($< 10^{-12}$ N) samt applicera både statiska och dynamiska krafter i biologiska system med utmärkt precision. Optisk pincett är därför en utmärkt teknik för mätning av intermolekylära krafter och för bestämning av biomekaniska strukturer och dess funktioner.

Vissa typer av bakterier har specifika vidhäftningsorganeller som kallas för pili. Dessa förmedlar vidhäftningen till värdceller och är därför viktiga vid bakteriens första kontakt. En djupare förståelse av pilis uppbyggnad och biomekanik kan därmed ge information, som kan vara vital i framtagandet av nya mediciner som förhindrar bakteriella infektioner. Detta är av stor vikt i skenet av den ökande antibiotikaresistensen i vårt samhälle.

I denna avhandling presenteras konstruktionen av en experimentell uppställning av kraftmätande optiskt pincett tillsammans med resultat från biomekaniska undersökningar av vidhäftande bakteriella organeller. Kraftmätande optisk pincett har använts för att karakterisera tre olika typer av pili, P, typ 1, och S pili, vilka kan uttryckas av uropatogena *Escherichia coli*. En kvantitativ biofysikalisk modell som beskriver deras förlängningsegenskaper under pålagd kraft har konstruerats. Modellen bygger på pilis strukturella uppbyggnad samt på dess respons som uppmäts med den kraftmätande optiska pincetten. Modellen beskriver de biomekaniska egenskaperna väl för alla tre pili. Dessutom kan ett antal specifika bindnings- och subenhetsparametrar bestämmas, t.ex. interaktionsenergier och bindningslängder. Skillnaden mellan dessa parametrar hos de tre pilis samt deras olika kraftrespons har jämförts. Detta arbete har dels resulterat i en förbättrad förståelse av pilis biomekaniska funktion och dels i en metod som, med hög noggrannhet, tillåter oss att bestämma ett antal biomekaniska egenskaper hos olika organeller på ett effektivt sätt.

CONTENTS

THIS THESIS IS BASED UPON
THE FOLLOWING APPENDED PUBLICATIONS

I. **M. Anderson**, E. Fällman, B.E. Uhlin, O. Axner "*A sticky chain model of the elongation of Escherichia coli P pili under strain*". Biophysical Journal 90 (2006) 1521-34.

II. **M. Andersson**, E. Fällman, B.E. Uhlin and O. Axner "*Dynamic force spectroscopy of the unfolding of P pili*". Biophysical Journal 91 (2006) 2717-2725.

III. **M. Andersson**, E. Fällman, B.E. Uhlin, O. Axner "*Force measuring optical tweezers system for long time measurements of Pili stability*". SPIE (2006) 6088-42.

IV. E. Fällman, S. Schedin, J. Jass, **M. Andersson**, B.E Uhlin and O. Axner "*Optical tweezers based force measurement system for quantitating binding interactions: system design and application for the study of bacterial adhesion*". Biosensors and Bioelectronics, 19(11) (2004) 1429-1437.

V. M. Klein, **M. Andersson**, O. Axner and E. Fällman "*A dual trap technique for reduction of low frequency noise in force measuring optical tweezers*". Applied Optics 46 (3): 405-412 JAN 20 2007.

VI. O. Björnham, O. Axner, **M. Andersson** "*Modeling of the elongation and retraction of Escherichia coli P pili under strain by Monte Carlo simulations*". European Biophysics Journal (2007) in press.

VII. **M. Andersson**, B.E. Uhlin, E. Fällman "*The biomechanical properties of E. coli pili for urinary tract attachment reflect the host environment*". Biophysical Journal 93 (2007) 3008-3014.

VIII. **M. Andersson**, O. Axner, F. Almqvist, B.E. Uhlin, E. Fällman "*Physical properties of biopolymers assessed by optical tweezers*". ChemPhysChem (2007) in press.

IX. **M. Andersson**, E. Fällman "*Characterization of S pili — investigation of their mechanical properties*". Submitted (2007).

ADDITIONAL WORK IN THE FIELD OF STUDY
NOT INCLUDED IN THE THESIS

X. E. Fällman, S. Schedin, **M. Andersson**, J. Jass, O. Axner "*Optical tweezers for the measurement of binding forces: system description and application for the study of E. coli adhesion*". SPIE 4962 (2003) 206-15.

XI. E. Fällman, **M. Andersson**, S. Schedin, J. Jass, B.E. Uhlin, O. Axner: "*Dynamic properties of bacterial pili measured by optical tweezers*". SPIE 5514 (2004) 763-33.

XII. E. Fällman, S. Schedin, J. Jass, **M. Andersson**, B.E. Uhlin, O. Axner "*Book of Abstracts of the 5th International Conference on Biological Physics ICBP2004*". Gothenburg, 2004, p. B07-279.

XIII. E. Fällman, **M. Andersson**, O. Axner "*Techniques for moveable traps influence of aberration in optical tweezers*". SPIE (2006) 6088-36.

XIV. **M. Andersson**, E. Fällman, B.E. Uhlin, O. Axner: "*Technique for determination of the number of PapA units in an E Coli P pilus*". SPIE (2006) 6088-38

XV. **M. Andersson**, O. Axner, B.E. Uhlin and E. Fällman "*Optical tweezers for single molecule force spectroscopy on bacterial adhesion organelles*". SPIE (2006) 6326-74.

XVI. **M. Andersson**, O. Axner, B.E. Uhlin, E. Fällman "*Characterization of the mechanical properties of fimbrial structures by optical tweezers in P. Hinterdorfer*". G. Schutz, P. Pohl (Eds.), VIII. Annual Linz Winter Workshop. Trauner, Linz, Austria, 2006, p. 19-22. ISBN:3-85499-163-0

# 1. INTRODUCTION

## 1.1. Light as a tool for trapping

The development of Optical Tweezers (OT) during the last years has opened a new exciting era within the field of bioscience. OT provide scientists with novel tools that allow small objects, e.g., cells and bacteria, to be trapped and manipulated on demand, non-invasively and non-destructively. In fact, OT have during the last decade, grown into a powerful technique for assessment of minute forces in the field of nanoscience in general and those in biological systems in particular. The possibility to probe weak forces involved in micro-biological systems, such as bacterial/cell adhesion, entropic and protein-folding forces, and the forces involved in stretching biological polymers, has improved the understanding of the intricate function of biological matter [1-8]. OT have indeed transformed the ordinary light microscope from a device for passive observation to a versatile tool for active manipulation and controlled measurement of biological objects.

Even though optical tweezers is a rather young technique, the basic physics behind trapping of particles by light has been known for centuries. In 1619, Johannes Kepler proposed that it is the radiation pressure that causes the tail of comets to always point away from the sun no matter where it is located during its journey. About 250 years later, James Clerk Maxwell came up with a theoretical model that proved what Kepler predicted: *"Light can exert forces to matter"* since momentum is transferred when an electromagnetic field interacts with a target. Although the radiation pressure in general is rather weak, it can be substantial for high intensities and small particles. For example, it is the opposing force to the sun's gravity, which keeps it from imploding.

However, light pressure had little practical use until the development of the laser. These intense light sources provided scientists with highly collimated light beams that could be focused to small spots and be used for manipulation of atoms and molecules. In the 1970s, the first demonstration of a counter-propagating laser trap was made by Arthur Ashkin [9]. He demonstrated that optical forces could move and levitate micrometer sized dielectric particles both in liquids and air [10]. Even though it was not possible at that time to move the trapped particle in the direction opposite to the propagation of the light it was the beginning of the single-beam gradient trap that today is known as optical tweezers.

A few years later, in 1986, Ashkin again demonstrated trapping of particles but this time it was the single-beam gradient trap showing its future potential. In comparison to the radiation pressure used in the counter-propagating trap it was now the spatial gradient force that pulled the dielectric particles towards the focus of the laser light. A year later, the first organic materials (bacteria, viruses and cells) were trapped with a green argon ion laser and a Nd:YAG laser [11]. This can be seen as when optical tweezers entered the field of Biophysics, showing the use and advantages of non-invasive micromanipulation. The possibility to trap organelles and chromosomes *in vivo*[1] without damaging the cells has made it possible to study cells and their interactions in a new way. Since then, a large number of applications have evolved into an exciting and fast growing field of science.

One of the more powerful and recent applications of gradient traps is force measuring optical tweezers (FMOT), which is a direct consequence of the fact that a trapped spherical object behaves as confined in a harmonic potential. As a consequence, a displacement of the trapped object from the equilibrium position in the trap leads to a restoring force that follows Hookes law, thus proportional to the displacement. This provides a possibility to apply and measure forces and displacements with sub-pN ($10^{-12}$ N) and sub-nm ($10^{-9}$ m) resolution [12, 13]. OT have thereby become a versatile and sophisticated tool both for micro-manipulation and sensitive force measurements.

The function of proteins is directly related to their structure [14]. Static structural pictures of proteins and polymers assessed by atomic force microscopy, nuclear resonance images, and crystallographic methods have helped scientists to gain knowledge of the architecture and function. In addition, the stability of proteins has been analyzed with calorimetry measurements, a method that provides binding enthalpies from average values of large ensembles [15]. In contrast, the FMOT technique allows for various types of dynamic studies on a single molecular level. This implies that individual molecular responses can be assessed, which brings in a new level of knowledge about biological systems and their responses. In addition, it is possible to map intermediate states in the energy landscape that would be experimentally unobservable by calorimetry methods [15]. These possibilities have increased the knowledge of the studied objects and helped scientists to build and experimentally verify theories of their biomechanical functioning. In conclusion, the advent of a new tool such as FMOT, which can probe the micro-machinery of macromolecules with high sensitivity and speed, has opened new means to study a variety of processes in biological systems and will for any foreseeable future continue to improve our basic knowledge of such systems.

---

[1] Inside the cell.

## 1.2. Applications of optical tweezers

The use of light to trap and move small objects has thus opened new doors in the microscopic world of manipulation in general and in sensitive force assessment in particular. Examples of applications of OT are; cell sorting, actively alteration of polymer structures (DNA melting, membrane deformation strength and unraveling of protein polymers), application of stall forces, characterization of molecular motors (myosin and kinesin), and measurement of binding forces in the biological and medical fields [16-22].

Precise manipulation / sorting techniques are important for many types of applications and the advantages of optical tweezers system in comparison with other mechanical manipulation techniques are many. The possibility of *in vitro*$^2$ manipulation with light is not only a sterile application but the biological effect is almost non-existing, e.g., the cells are not affected in a way that will inhibit their growth [23-25]. In addition, the development of non-contact micro dissection tools for pathology, known as laser scalpels, has complemented optical tweezers as a useful technique in bioscience [26]. OT are easily implemented into microscopes since most of the necessary optics are already there for imaging purposes. Actually, implementing OT in an existing microscope is more or less straightforward and requires only a minimum number of components and alterations. It can therefore be seen as quite an inexpensive investment, at least if moderate trap strengths are needed.

Mechanical properties of biopolymers can be assessed by various single molecule force spectroscopy methods that spans the force regime from sub-pN (FMOT, magnetic tweezers) up to several nN (AFM, BFM, microneedles) [27-29]. Many types of non-covalent interactions, which play an important role in biological systems, are of the orders of a few tens of pN. The most common force measuring techniques are AFM, mainly since it is commercially available, and FMOT, primary due to its high sensitivity. The main advantages of FMOT are its high sensitivity and high flexibility. Due to its construction, FMOT can measure forces that are approximately one order of magnitude lower than those of AFM (sub pN *vs.* low pN forces). Another advantage of FMOT with respect to AFM is the possibility to make a fast change of the force transducer. In an AFM system, the tip must be replaced if contaminated whereas a typical sample in a FMOT consists of numerous beads that all can be quickly replaced, calibrated and used as transducers.

Since many types of adhesion and entropic forces in biological systems (receptor-ligand bonds and entropic forces in large chain-like biomolecules) are often in the low pN range, FMOT is a particular suitable tool for assessments of such forces [30]. In addition, FMOT are also suitable for measurements of static forces (e.g., stall forces of motor proteins) as well as dynamic properties of biomolecules and interactions, even those originating from a single bond [31, 32]. A variety of experiments associated with folding of macromolecules and specific adhesion have therefore been performed during the last years [30-33]. These studies have revealed new knowledge about mechanical or biophysical properties of various biopolymers and given new insight into the behavior of biological macromolecules.

---
$^2$ In a laboratory environment.

## 1.3. Specific aim of this thesis

This thesis reflects the work — experimental as well as theoretical, most of it performed within the inter-disciplinary field of Biological physics — that I have pursued during the years as a PhD student. The experimental section describes the constructing of a state-of-the-art optical tweezers setup used mainly for force measuring applications of biopolymers and specific receptor-ligand interactions. The work involves development of a measuring model system for assessment of forces in bacterial adhesion organelles. The work has also included the construction of a tailor made computer program used for controlling the equipment and for data acquisition. In addition, a biomechanical model of the force-extension properties of adhesion organelles has been developed.

The scientific work has been aimed at studies of adhesion organelles expressed by Uropathogenic *Escherichia coli* (UPEC) bacteria. These organelles are responsible for the initial contact with host cells and they serve therefore as an important virulence factor. Moreover, this work is partly motivated by the fact that the heavy use of antibiotics during the last decades has increased the bacterial resistance towards such drugs. Consequently, new drugs or ways to prevent bacterial infection are urgently needed. My work has aimed towards characterization of such organelles, with the specific goal of understanding their intricate intrinsic biomechanical properties. This work will hopefully help clarifying questions regarding the working principles of pili and their role in the adhesive process, in particular the ability to resist the cleaning action of the urine flow, but also issues regarding similarities between different pili, why they show differences, and if those can be related to natural environments. Such questions would be difficult to assess with other methods.

## 2. THE BASICS OF OPTICAL TWEEZERS

The principle of optical tweezers is built upon the fact that a photon carries a momentum, $p = h/\lambda$, where $h$ is Planck's constant and $\lambda$ the wavelength of the light, that is changed during interaction with matter through refraction or scattering. The change in momentum implies that a force acts on the object. Such a force can, for convenience, be decomposed into a scattering- and a gradient force, where the former acts in the direction of the propagation of the light and the latter in direction of the gradient of the intensity. Moreover, a particle is trapped at the point where the gradient and scattering forces are in balance, i.e., slightly after the focus. For a single beam optical trap this requires a steep intensity gradient, which most conveniently is achieved by the use of a high numerical aperture (N.A.) objective, as is illustrated in Fig. 1.

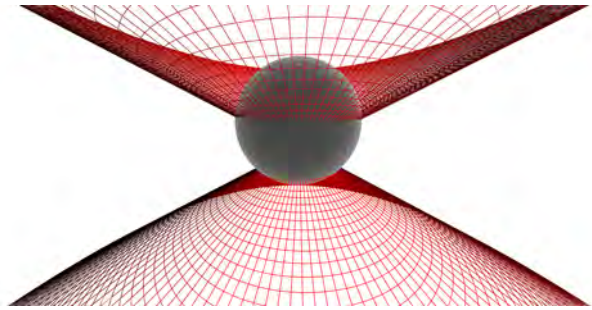

Figure 1. An illustration of a bead trapped in a focused laser beam. The bead is being pulled to the position in the focal region where the forces balance out.

Although the concept of radiation pressure has been known for centuries, it is not trivial to provide a theoretical description of micron sized optical trapping by a strongly focused laser beam that is both generally valid and user friendly. To simplify the treatment of optical trapping two size dependent approximations are made that agrees well with experimental findings, referred to as the *Rayleigh* and the *ray optics* approximations. The former is valid for objects much smaller than the wavelength of the light, i.e., $\ll \lambda$, whereas the latter is valid for larger objects, hence $\gg \lambda$. Ashkin derived a theory of the optical forces produced by light interacting with micron sized objects based upon the ray optics approximation [34, 35]. This picture has also been used in this thesis to describe the principle of optical trapping. Therefore, the Rayleigh approximation will only shortly be introduced, while a more detailed review of the ray optics theory will be given. The ray optics approach has also the advantage that it describes optical trapping in an intuitive manner.

The models presented are important when it comes to providing a basic description of the forces in the system. They will not, however, provide a complete picture of the forces in an optical trap. Problems regarding the geometry of the trapped object, the amount of light in the sample plane, light-induced heating, and aberrations from the optical system make precise calculations difficult. It is therefore important to use well established empirical methods to assess the strength of the optical trap when it is used as a force transducer [36]. This section describes, in short, the basic principle of optical trapping, optimization, detection and principle of calibration of force measuring optical tweezers.

## 2.1. The theory behind the forces in an optical trap

### 2.1.1. The Rayleigh regime

In the Rayleigh regime the object is described as a dielectric dipole (i.e., as a point) whereas the light is described as a non-homogeneous electromagnetic field [37]. Whenever there is a gradient in the electromagnetic field, there will be a force induced on the particle proportional to the gradient of the field and thereby also proportional to the gradient of the intensity of the light. A highly focused Gaussian beam has an intensity maximum in its centre, why it will give rise to a three-dimensional gradient of the laser light, which in turn produces a force directed towards the centre. The gradient force (dipole force), $F_g$, is then induced by the field and described as,

$$F_g = \frac{2\pi r^3 n_1}{c} \left( \frac{m^2 - 1}{m^2 + 2} \right) \nabla I, \qquad (1)$$

where $c$ is the speed of light, $r$ is the radius of the object, $m$ the effective index of refraction $(n_p / n_1)$, where $n_p$ and $n_1$ are the index of refraction of the particle and medium, respectively, and $I$ the intensity of the light. Equation (1) shows that the gradient force is proportional to the volume of the particle (which originates from the fact that the number of polarizable molecules in the object scales linear with its volume), and to the gradient of the intensity of the trapping light, i.e., $F_g \propto \nabla I$. The scattering force $F_s$ is directly proportional to the intensity of the light and is given by,

$$F_{sc} = \frac{128\pi^5 r^6 n_1}{3\lambda^4 c} \left( \frac{m^2 - 1}{m^2 + 2} \right) I. \qquad (2)$$

These expressions show that both the gradient and scattering forces are strongly influenced by the size of the particle. As the particle becomes larger, the scattering force is dominating, which makes optical trapping unstable.

## 2.1.2. The ray optics regime

For objects larger than the wavelength of the light, i.e., $d \gg \lambda$, ray optics can be used to describe the forces acting on a spherical bead [38]. In short, a beam of light is decomposed into a bundle of rays, where each ray has a certain power. A ray is represented as a straight line that follows Snell's law of refraction and Fresnel's equations for reflection and transmission as schematically illustrated in Fig. 2. When a ray is refracted parts of its momentum is transferred from the ray to the particle. A summation of the change of momentum from all individual rays gives the total net force acting on the particle.

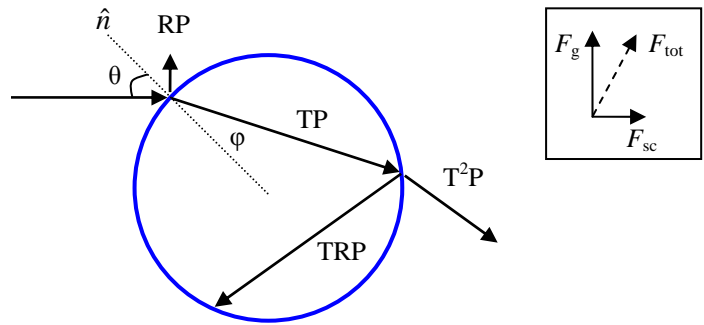

Figure 2. A ray diagram of the power contribution from a single ray, P, propagating through a spherical particle with an index of refraction larger than the surrounding medium. The ray undergoes both reflection, R, and transmission, T, at all surfaces. $\hat{n}$ denotes the normal to the surface of the sphere. The rightmost insert represents the total force acting on the particle from the single ray due to contributions from the scattering and gradient force.

For lateral trapping, i.e., perpendicular to the propagation of light illustrated in Fig. 3A, the gradient force arises from the non-uniform intensity of the laser beam (e.g., a Gaussian distribution), which creates a restoring force in the direction of highest intensity. Moreover, the scattering force, which comes from the reflection at the surface of the particle, propels the particle forward and counteracts axial trapping. Ashkin solved this problem in the beginning by using counter-propagating lasers that balanced the object through the scattering force [10]. However, the single gradient optical tweezers technique solves this problem by producing a strong gradient force in the opposite direction of the scattering force. This requires a large gradient in the light intensity in the axial direction, which can be obtained by a strong focusing of the light, as is illustrated in Fig. 3B.

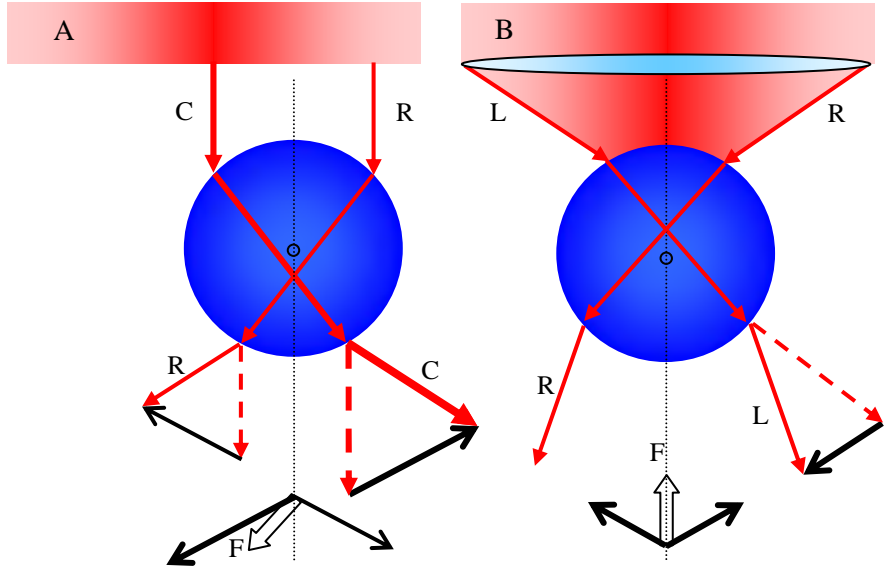

Figure 3. Illustration of a sphere in a Gaussian shaped laser beam where the trapping force is described by the ray optics theory. In A) two rays, C with high intensity and R with low intensity, are both refracted by the sphere. The momentum change of the refracted rays, illustrated by the solid rays, affects the sphere with an opposite force. However, since C has a higher power than R, the bead will experience a force to the left. Thus, in A the Gaussian intensity distribution of the beam directs the sphere towards the centre of the beam. In B) a high N.A. objective focuses the rays and the change of momentum creates a restoring force towards the focus in the axial direction, i.e., where the rays should meet if the sphere did not refract them.

For a single ray Ashkin showed that the scattering force, $F_{sc}$, and the gradient force, $F_g$, can be written as,

$$F_{sc} = \frac{n_1 P}{c}\left(1 + R\cos 2\theta - T^2 \frac{\cos(2\theta - 2\varphi) + R\cos 2\theta}{1 + R^2 + 2R\cos 2\varphi}\right), \quad (3)$$

$$F_g = \frac{n_1 P}{c}\left(R\sin 2\theta - T^2 \frac{\sin(2\theta - 2\varphi) + R\sin 2\theta}{1 + R^2 + 2R\sin 2\varphi}\right), \quad (4)$$

where P is the power of the ray, $\theta$ is the angle of the incident ray and $\varphi$ is the angle of the refracted ray [38]. The first term, $n_1 P/c$, is the momentum per second transferred by the ray. R and T are the Fresnel reflection and transmission coefficients, which give the fraction of light being reflected or transmitted at an interaction. The total force delivered by a beam to a spherical particle is then found by summing the contributions of all rays using Eqs (3) and (4). It is thereby possible to calculate dimensionless efficiency factors, $Q_{sc}$ and $Q_g$, for the entire beam that depends on the laser wavelength, N.A., its polarization, the laser mode, the geometry of particle and the relative index of refraction, given by $F_{sc} = Q_{sc} n_1 P / c$ and $F_g = Q_g n_1 P / c$. It is also possible to express the maximum force generated by the trap as

$$F = Q_{tot}\left(\frac{nP}{c}\right), \tag{5}$$

where $Q_{tot} = \sqrt{Q_{sc}^2 + Q_g^2}$. The strength of the trapping force is thereby controlled by the laser power, the refractive index of the surrounding medium and the $Q$-value [36]. Numerical calculations of the trapping forces for various conditions are given in [38]. As an example, the maximum gradient force of a polyester sphere in water is achieved for a beam with marginal rays at angles of ~70°. Under these conditions a restoring force of ~200 pN is reached for 150 mW of laser power.

### 2.1.3. Optimizing trapping

As is described above there are a few parameters that allow us to optimize the trapping strength of single gradient optical tweezers. The easiest and most straightforward is to increase the laser power since this determines the flux of photons in the focal region and thereby the trapping power. However, high laser light intensities can cause damage of the optical components, especially the microscope objective, which mostly are constructed for high transmission of visible wavelength [36]. Moreover, local heating of the medium from absorption of light, which is of the order of typically a few K/Watt, can change the conditions for the biological object under study as well as the viscosity of the liquid and thus influence the trap stiffness [39, 40]. In addition, trapping of organic material with high intensities can also lead to photo damage [11, 41]. Therefore, low laser power, tens of mW should be use for manipulation of biological objects, e.g., cells whereas moderate power, a few hundreds of mW can be used during force measurements since the probe transducers are often made of non-organic material, e.g., polystyrene. Therefore, even though high power lasers can be bought quite cheaply the maximum amount of light should be restricted to a few watts.

It is most often not possible to optimize the index of refraction for force measurements of biological samples by modifying or changing the solution. First, since protein interactions are strongly affected by pH and salinity the object under study requires a solvent that represent the natural environment for trustworthy results. Secondly, cells are negatively affected if placed in an inappropriate environment. Instead, the strength of the trap is best optimized by changing the Q-values of the trap.

Since the trap strength relies on the gradient of the focused laser light, a tight focus will trap better than a wide. In addition, light of short wavelength produces a smaller focus as compared to light with long wavelength due to diffraction. This suggests that the best conditions should be obtained by using short wavelength laser light. In reality, however, this is strongly restricted by the large light absorption coefficient of the sample under study in this wavelength region. Trapping biological samples non-intrusively should therefore not be performed with lasers of too short wavelength [36]. Consequently, trapping of biological objects is preferably done by light in the near infra red region.

It was discussed above that the maximum gradient force is achieved with highly converging rays. This is realized in practice by the use of a high numerical aperture objective (often with N.A. > 1.2). The maximum angle of incident is defined as,

$$\text{N.A.} = n_1 \sin(\theta),\qquad(6)$$

where $n_1$ is the index of refraction of the immersion medium (oil or water) and $\theta$ is the half angular aperture. An illustration of the distribution of rays refracted by a spherical object created by two objectives with dissimilar N.A. is shown in Fig. 4. It is seen from the second panel, which has a N.A. ~1, that the momentum change of the photons is larger than for the left image which has a lower N.A. Thus, the trap created by a high N.A. objective will produce a stronger axial restoring force.

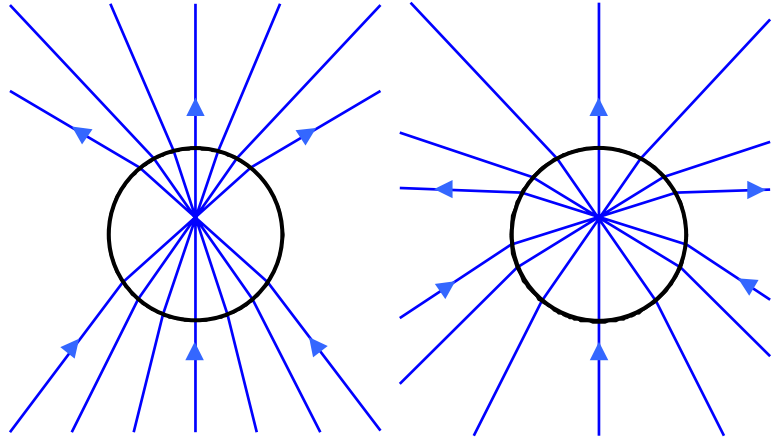

Figure 4. A geometrical description of the refraction of light through a spherical particle, simulated with a raytracing program, from an objective with low and high numerical aperture.

Moreover, as is illustrated in Fig. 5, the outermost rays, i.e., the ones with the largest angle, are those that correspond to the rays that strike the rim of the entrance pupil. In order for the outermost rays to have any appreciable trapping power in the axial direction, the entrance pupil should be overfilled with light. Typically, ~80 % of a Gaussian shaped beam should be used to achieve good axial trapping [42].

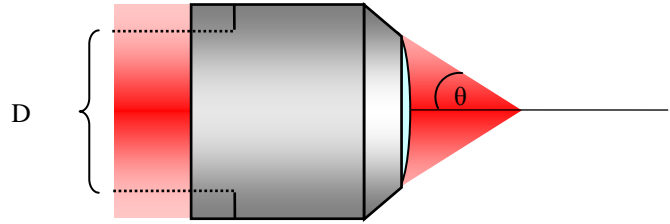

Figure 5. The entrance pupil, D, of the objective should be overfilled for strong axial trapping. Thus, the overfilling provides the focused beam to have more powerful marginal rays.

## 2.2. Force detection techniques

Optical tweezers are not only capable of trapping objects for manipulation purposes; they can also be used directly to measure and apply forces in a delicate and sensitive manner. The optical trap confines the particle in a three dimensional harmonic potential, which for a single direction can be described as $E(x) = kx^2/2$, where $k$ is the harmonic constant. Thus, the force is linearly dependent on the displacement and follows a Hookean behavior, as illustrated in Fig. 6. Thus, by measuring the position of a bead in a trap, the force to which it is exposed can accurately be determined.

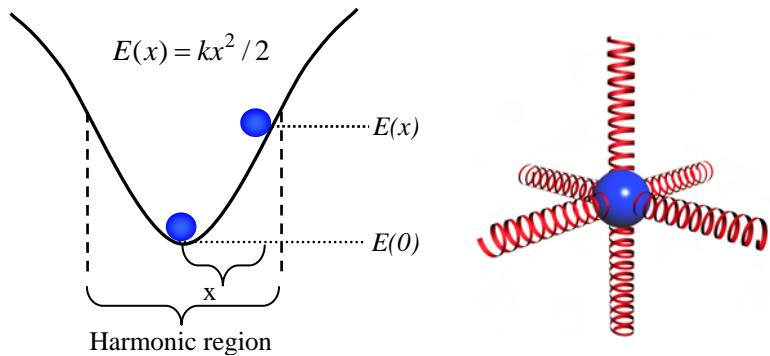

Figure 6. The left illustration shows a trapped particle confined in a harmonic potential with a trapping constant of $k$, for small displacement. The particle can be envisioned as if it would be attached to microscopic springs, right illustration. The energy of the particle is then given by $E = k_i x_i^2 / 2$, where $i =$ x, y, z directions.

### 2.2.1. Position detection techniques

A number of techniques for probing the position of a confined bead have been developed over the years [40, 43]. The various techniques are based on slightly different principles and have therefore dissimilar properties. One direct approach uses a video camera that together with a centroid-finding algorithm can determine the position of the object. However, the accuracy and speed of this technique is limited by the optical resolution and the maximum sampling speed, which is restricted to the image acquisition rate, typically up to ~120 Hz [40].

A better approach is therefore to use a laser-based position method, either by employing the laser that is used for trapping or a second low power probe laser. One sensitive laser monitoring technique is based upon polarization interferometry using the trapping laser [36]. The drawback of this method is that it can only measure displacements in one dimension. To circumvent such problem, back focal plane detection (BFP) can allow for measurements of all axes of interest. It relies on the interference between the forward scattered light from the bead and the unscattered light [44-46]. The interference is monitored with a position sensitive detector (PSD) or quadrant photodiode (QPD) on the optical axis at a plane that is conjugate to the focal plane of the condenser [44].

Finally, the implementation of a separate low power probe laser provides the same degrees of freedom as the BFP detection but with additional advantages [43, 47, 48]. The last method has been implemented in the work presented in this thesis. A more detailed description is included below.

### 2.2.2. The probe laser technique

In the probe laser technique, the light is focused slightly below the force measuring bead. The bead acts as a lens, which collects and focuses the light, that together with the microscope condenser creates an optical system that images the light onto a detector, in our setup a PSD. The use of a separate probe laser gives full control of the focus, i.e., it can be freely altered with respect to that of the trapped object. This makes it possible to optimize the linearity and the strength of the probe response separately from the properties of the trap. Furthermore, it is easy to match the probe laser wavelength to a detector, which can be difficult with a laser running in the IR-region. For example, it has been shown that light from an Nd:YAG laser that operates in IR can give rise to low pass filtering effects on QPDs made of silicon [49]. If needed to, the probe laser technique makes it possible to monitor multiple objects. In addition, the probe beam does not require the same high numerical aperture as the trapping beam (which has to impinge onto the bead with large incidence angles to provide adequate trapping); it suffices if the probe beam is focused by the trapped bead, as illustrated in Fig. 7. This implies that the probe beam is less influenced by aberrations, since only the inner part of the objective lenses is used, and that a dry condenser can be used instead of an oil condenser, since the probe light is less focused. Figure 7 shows schematically how such an experiment can be performed. The wired frame illustrates the trapping light whereas the solid represents the probe light. The trapped bead and the condenser optics create an optical system that images the light onto a detector. In Panel A the bead is in equilibrium and the probe spot is centred on the PSD. In Panel B, the separation of the bacteria and the trapped bead creates an external force that shifts the position of the bead in the trap. The probe light is deflected and the spot on the detector is moved from the centre position on the PSD. Thus, if the detector signal is calibrated it is possible to convert the signal to a displacement.

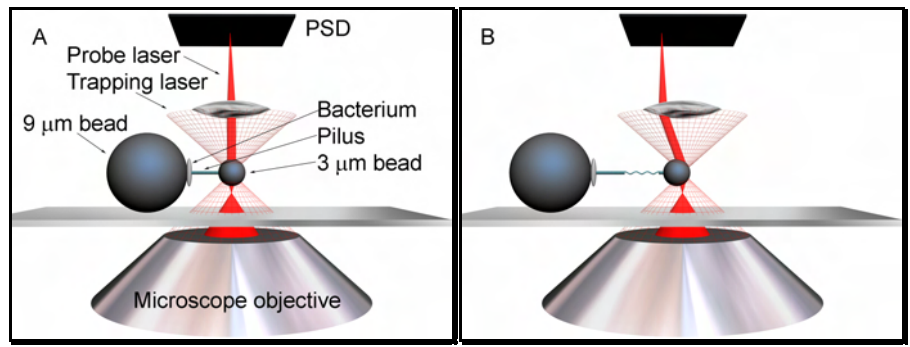

Figure 7. A schematic illustration of a trapped bead used as a force transducer and probed by a laser light source (solid red). The light is monitored on a PSD that through a calibration procedure converts the deflection to a force. Panel A shows the probe light focused on to the detector by a bead in its equilibrium position, i.e., when no external force is applied, whereas panel B shows the same bead off axis, causing a deflection of the light.

## 2.3. Force calibration techniques

As mention previously, the Rayleigh and ray optics theories give descent estimates of the trap strength for small and large particles as compared to the wavelength. However, most trapped objects are of the order of the laser wavelength, ~1 µm. Although there are complex theories that describe the axial force for such range, they do not give the complete picture [50]. The theories presented above are not accurate enough when absolute force measurements are needed. Consequently, several empirical techniques have been developed to calibrate the optical forces [51]. These techniques have different approaches and advantages, which implies that it is important to use them with care.

The empirical techniques are based on probing the position shift of an object on a detector. However, the signal from the detector is not providing information of the absolute displacement of the bead. Therefore, an absolute force measurement requires two types of calibration. First, the relation between the displacement of the bead and the detector signal must be determined, i.e., a conversion from voltage into displacement units. This step could be performed in different ways, either by fixating the bead to the cover slide and simply move the stage with known steps and thereby correlate the movement of the stage to the detector signal. Such movement is possible since the stage often have nm resolution. However, regular trapping is often performed a few micrometer away from the cover slide, wherefore such calibration procedure can lead to an incorrect estimation of the conversion factor. Alternatively, the detector signal is correlated by moving the trap, while keeping the probe laser fixed. This procedure requires that the movement of the trap is calibrated with an absolute scale, which is described in chapter 4.

The second calibration converts the displacement of a trapped bead to a force. However, to convert the detector response to a force, the object must be calibrated against a known force. There are several ways of calibrating the stiffness of a trap, where a reliable and fast method is the one based upon power spectrum analysis [36, 51]. The method is based upon the fact that a trapped bead explores the potential through thermal agitation (Brownian motion). However, the optical potential confines the particle and suppresses the motion, mainly the large fluctuations. These fluctuations, provides a statistical approach to correlate the displacement of the bead to a known force. Figure 8 shows a two dimensional histogram of the lateral position distribution of a 3.0 μm bead in an optical trap.

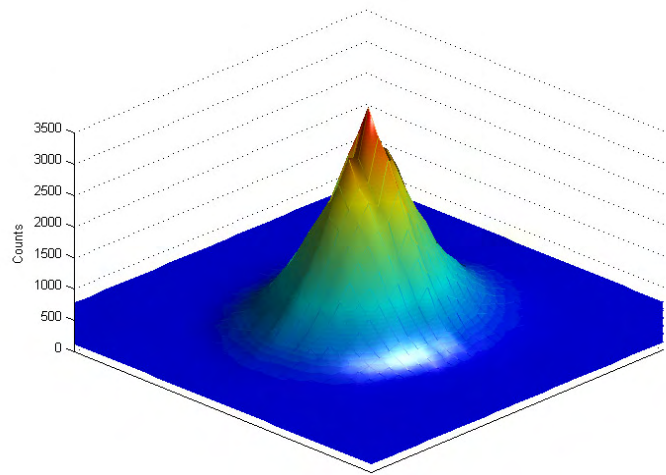

Figure 8. A two dimensional histogram plot of the accessed lateral positions explored by a trapped 3 μm bead. The data has been subjected to a smoothening surface plot for representative purpose.

The motion of a trapped bead can be described by the Langevin equation extended with a harmonic part as

$$\gamma \frac{dx}{dt} + kx = F(t), \qquad (7)$$

where γ is the viscous drag coefficient and $F(t)$ the random force, which averages out to 0 [52]. The mass and moment of inertia of the bead can be neglected since such a small particle is strongly overdamped in the liquid [53]. For a particle confined in a trap, the displacement fluctuations in the frequency domain can be derived through a Fourier transform of Eq. (7). The power spectrum of $F(t)$, $S_F(F)$, can, for an ideal Brownian motion, be written as, $S_F(F) = |F(f)|^2 = 4\gamma kT$, where $F(f)$ is the Fourier transform of $F(t)$ [53]. The power spectral density of the motion of a particle, $S_x(f) = |X(f)|^2$, has a Lorentzian shape and can be expressed as

$$S_x(f) = \frac{kT}{\gamma \pi^2 (f^2 + f_c^2)}, \qquad (8)$$

where $k$ is the Boltzmann constant, $T$ the temperature, and $f_c$ is the so called corner frequency, which is related to the stiffness of the trap $\kappa$, as $f_c = \kappa / 2\pi\gamma$. The corner frequency divides the spectrum into two different regions. For frequencies $f \ll f_c$ the power spectrum is constant indicating a suppressed motion of the particle. On the other hand, for frequencies $f \gg f_c$ the power spectrum drops as $1/f^2$ which is characteristic for free diffusion. For these frequencies, the particle is not affected by the optical trap. A schematic illustration of Eq. (8) (solid curve) is shown in Fig. 9, where the two asymptotes (horizontal and tilted dashed lines) of the two regimes are visible. The intercept (vertical dashed line) indicates the corner frequency.

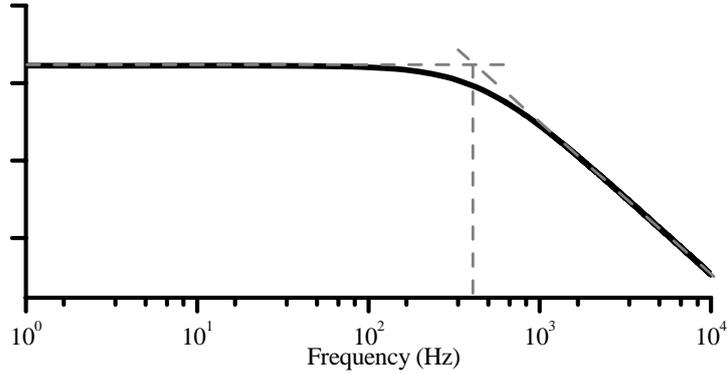

Figure 9. A schematic illustration of Eq. (8) shows a Lorentzian line shape. The corner frequency corresponds to the intercept of the two asymptotes and provides information about the stiffness of the trap.

Another common calibration method is based upon the Stokes drag force. In this technique, a known fluid flow is generated either by a syringe pump or by moving the sample stage [36]. A trapped particle will then be pushed away from the equilibrium position because of the drag. A spherical particle with radius $r$, in a flow with velocity $v$, far away from a surface, will experience a viscous drag force given by,

$$F(v) = 6\pi\eta rv, \quad (9)$$

where $\eta$ is the viscosity of the liquid. In practice, it is then possible to oscillate the sample stage with a given frequency and amplitude while holding the bead stationary. However, in most practical cases, Eq. (9) should be corrected with a sphere-surface separation factor according to [54]. In such case, the corrected Stokes drag force, $F'(v)$, can be expressed as,

$$F'(v) = \frac{6\pi\eta rv}{1 - \frac{9}{16}\left(\frac{r}{h}\right) + \frac{1}{8}\left(\frac{r}{h}\right)^3 - \frac{45}{256}\left(\frac{r}{h}\right)^4 - \frac{1}{16}\left(\frac{r}{h}\right)^5}, \quad (10)$$

where $h$ is the distance from the surface to the centre of the sphere. Figure 10 illustrates the Stokes drag force, from the Eqs (9) and (10), of a 1.5 µm particle (radius), 5 µm above the surface, for velocities of 5, 10, 20, 40 80, and 160 µm/s. As seen in the figure, the drag is ~2.5 pN for a velocity of 80 µm/s for the corrected equation. Consequently, measurements performed at high velocities, e.g., dynamic force spectroscopy measurements, should always proceed by a calibration carefully corrected for the drag.

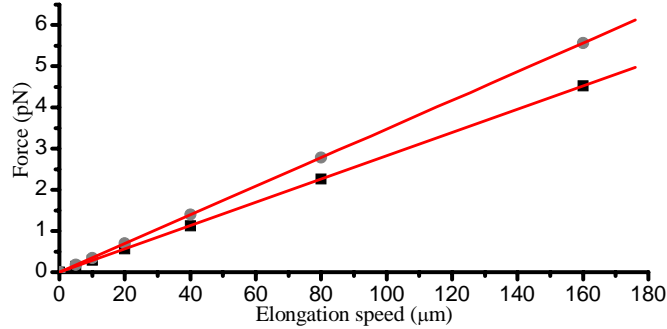

Figure 10. A schematic illustration of Eqs (9) and (10). The deviation between the infinite and wall corrected drag force is ~ 0.5 pN for a 1.5 µm (radius) particle at an extension speed of 80 µm/s.

# 3. CONSTRUCTION OF OPTICAL TWEEZERS

The construction of an optical tweezers system for trapping of particles is most often straightforward work. In principle, only a few optical components are needed, predominantly a laser for trapping, optics for beam shaping, optical mechanics for beam steering and a high numerical aperture objective to achieve a high spatial light gradient. Such a system can nowadays be constructed without too large cost [55]. Although not necessary, a slightly modified commercial microscope is useful, not only for imaging purposes but also to make the system user friendly. On the other hand, one should notice that this basic setup is restricted to simple manipulation and will not offer more than a trap and release function.

Optical force measurements are much more complicated and require further refinements of the experimental setup and the surrounding environment. First of all, additional components need to be implemented to the existing setup; in some cases a supplementary probe laser (depending on the method of detecting the position of the trapped object) and a detector system (including; filters, optics, mirrors, amplifiers, and detector electronics) for monitoring of the position of the trapped bead. The optical table that holds all components should be damped to minimize vibrations that otherwise would limit the measurements. Besides this, and to reach even better measurement conditions, the room should be temperature controlled, noise isolated, and air turbulence proof. Figure 11 shows a picture of undersigned's laboratory in which these issues have been implemented. In addition, high resolution steering of laser beams and samples are needed, which often requires piezo-controlled motors that need to be accurately controlled by dedicated programs. Moreover, a well tuned data acquisition system for calibration routines and position / force sampling must be implemented. Consequently, the inexpensive and easy assembled trap has become an expensive and elaborate construction, which on the other hand allows for highly sensitive force measurements. The following chapter describes the components, the principle of beam steering and detection technique used in the force measuring optical tweezers setup that undersigned has constructed.

The optical tweezers system constructed in undersigned's laboratory consists of an inverted microscope where a cw Nd:YVO$_4$ laser is used for trapping, whereas a fiber-coupled HeNe laser is used for probing the position of the trapped object. The light from the probe laser is monitored on a PSD detector mounted on top of the microscope. An additional piezo-controlled stage is mounted for precise steering of the sample under investigation. The instrumentation is schematically illustrated in Fig. 12.

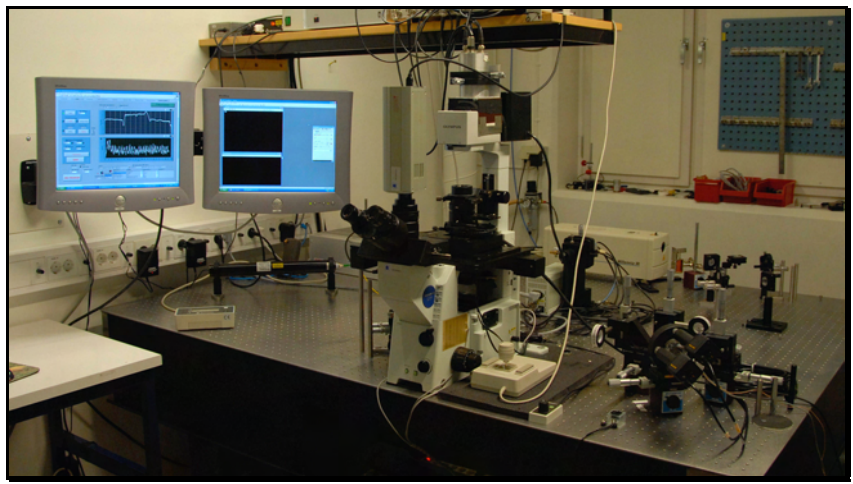

Figure 11. A picture of undersigned's optical tweezers setup which is build on an air floating optical table to reduce the influences of vibrations. A CCD camera is mounted on the microscope to allow high resolution imaging of the sample displayed on the right screen, whereas real time data acquisition of the sample under study is shown on the left screen.

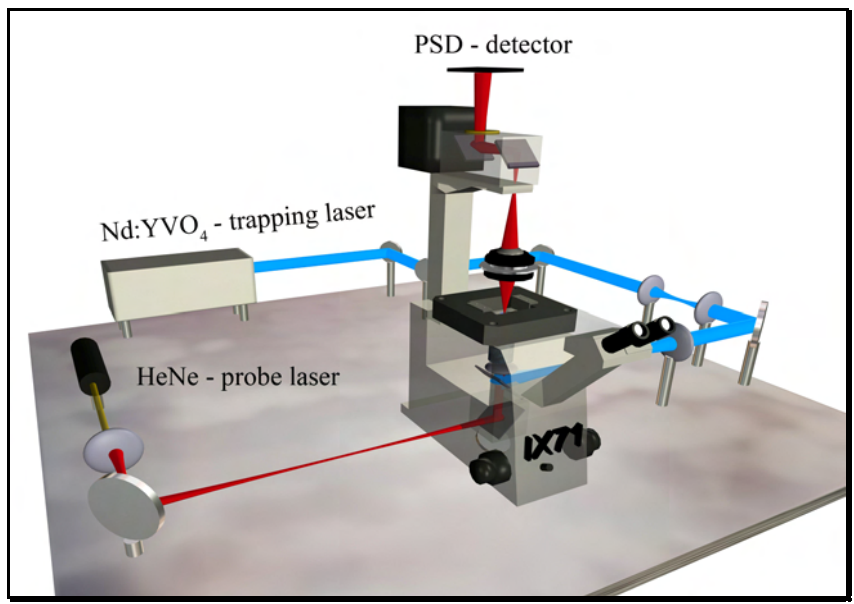

Figure 12. A schematic illustration of the optical tweezers system constructed by undersigned. The inverted microscope is modified for introducing laser light for trapping and probing (trapping, Nd:YVO$_4$, blue and probing, HeNe, red). The probing laser is fiber-coupled to reduce vibrations and provide easy alignment. The trapping light is blocked by laser line filters so that only light from the probe laser reaches the PSD-detector.

## 3.1. The microscope system

### 3.1.1. The microscope

The force measuring optical tweezers system was built on a vibration isolated (air floating) optical table[3] (TMC) around an inverted microscope[3] (IX-71, Olympus) modified for introducing laser light both for trapping and position monitoring of a trapped object. Moreover, for studies of specimens that require a local environment of 37° C a plastic chamber (incubation box) was constructed, which fitted the microscope and kept a constant temperature by applying heat to the body of the microscope and through a specially designed temperature regulating sample holder, not shown in the picture. The microscope was thermally isolated from the optical table by a 5 mm thick layer of bakelite.

### 3.1.2. Imaging of the sample

The development of digital cameras and advances in image processing techniques has increased the amount of accessible information in optical microscopy. High resolution charge-coupled device (CCD) cameras are able to capture the weak light from fluorescence or record the dynamic events of diffusion processes with multiple frames. Therefore, a CCD cooled colour camera[3] that allows both high resolution imaging and video recording was mounted on the microscope. In addition, together with digital imaging computer toolboxes it was possible to enhance the image quality and optimize the calibration procedure, as described below. Also, the camera provides higher accuracy during mounting procedures and surveillance during computer controlled measurements. A digital image of a typical sample is shown in Fig. 13.

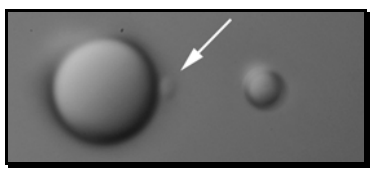

Figure 13. An image of a typical sample. The sample consists of functionalized 9 µm beads immobilized to the cover slide, 3 µm beads used as force transducers, and mounted bacteria (indicated by the white arrow).

---

[3] Information about various individual components is given in Table 1, which is located at the end of this chapter.

## 3.2. The optical tweezers system

### 3.2.1. Instrumentation

As is illustrated in Fig. 14, a linearly polarized diode pumped Nd:YVO$_4$ laser with a TEM$_{00}$ mode (Gaussian beam profile) was used for trapping. This laser provides good beam quality with low amounts of intensity and pointing fluctuations. In addition, its wavelength, 1064 nm, lies in a wavelength region where cells have a low absorption coefficient [41].

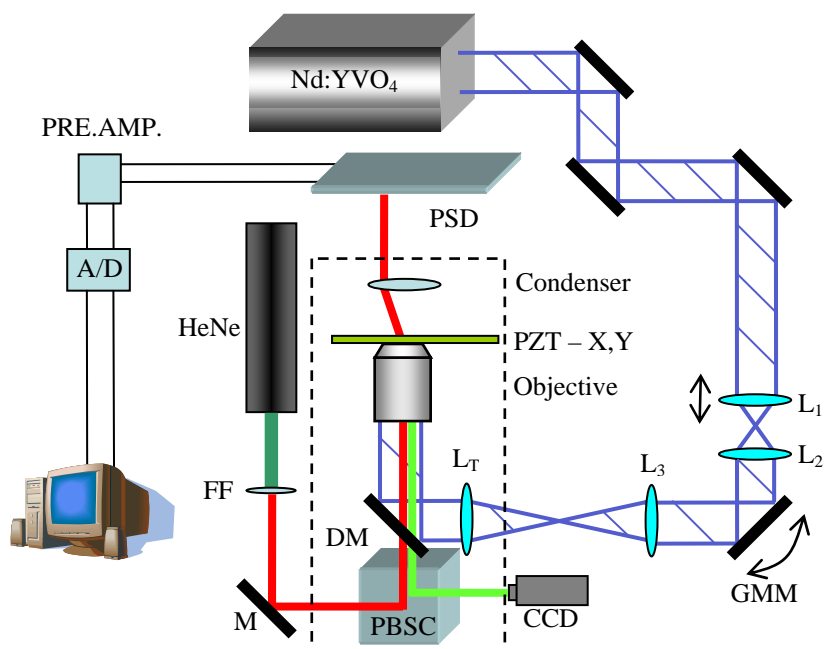

Figure 14. A schematic illustration of the FMOT setup where the dashed rectangular box represents the objects that are parts of the microscope. A cw Nd:YVO$_4$ laser is used for trapping. The z-direction of the trap inside the sample is controlled via the lens $L_1{}^3$ which together with lens $L_2{}^3$ (with the same focal length as $L_1$) forms an afocal lens system. A computer controlled gimbal mounted mirror, GMM[3], controls the lateral direction of the position of the trap in the sample plane. The entrance pupil of the objective is imaged onto the surface of the GMM by a second afocal lens system, produced by an external lens, $L_3{}^3$, and the tube lens $L_T$ inside the microscope. The light is introduced via a dichroic mirror, DM[3], coated for reflecting 1064 nm light and with a high transmission for visible light. The probe light from a fiber-coupled[3] HeNe laser with a fiber focuser[3], FF, is reflected via a mirror, M, and introduced via a polarizing beam splitter cube, PBSC[3]. The probe laser light is imaged via the microscope condenser onto a position sensitive detector, PSD[3]. The detector signal is amplified and filtered with a preamplifier[3] and collected via an A/D card[3] in a computer.

The trapping light was fed into the microscope through a set of optical components that shaped the beam to the correct size and divergence and provided a possibility for accurate adjustments of the trap in the lateral and axial plane. The principle of how to adjust the position of the trap in the sample is described in the next chapter. The laser was directed into the microscope by the use of an IR antireflection coated dichroic mirror[3] that was placed in the right side port of the microscope, which is normally used for imaging purposes. To ensure that the laser beam slightly overfilled the entrance pupil to ~80% of the high numerical aperture objective[3], a dummy objective was used together with a power meter. The dummy objective has the same entrance aperture as a regular objective except that it lacks lenses. It is thereby possible to change the width of the laser beam and measure the power before and after the objective until an appropriate overfilling is achieved. To avoid laser light reflections to the eyes, back scattering from all reflective surfaces, a laser blocking filter[3] was placed in front of the oculars. The laser was typically run at a power of ~1.0 W during force measurements, which gave rise to a stiffness of ~$1.5 \cdot 10^{-4}$ N/m (150 pN/µm).

### 3.2.2. Beam steering

As illustrated in Fig. 15, axial steering of the trap can be controlled by changing the divergence of the beam via an afocal lens system, $L_1$-$L_2$ [56]. A movement of lens $L_1$ a distance $\Delta d_{12}$ (where $\Delta d_{12}$ is a small change of the distance between $L_1$ and $L_2$) results in a corresponding axial displacement of the focus $\Delta z$ that is given by,

$$\Delta z = \left(\frac{f_{obj}}{f_T}\right)^2 \left(\frac{f_3}{f_2}\right)^2 \Delta d_{12}. \qquad (11)$$

where $f_2, f_3, f_T,$ and $f_{obj}$ are the focal lengths of lenses $L_2$, $L_3$, $L_T$, and $L_{obj}$, respectively. In this setup, the focal lengths of the lenses were chosen to 60, 150, 180, and 1.8 mm, which according to Eq. (11) results in a 6 µm trap displacement for a 10 mm lens movement. The high resolution of the micro meter positioning stage, < 0.5 µm, therefore allows for fine adjustment (~nm) of the focus (and thereby the trapped object) in the axial direction.

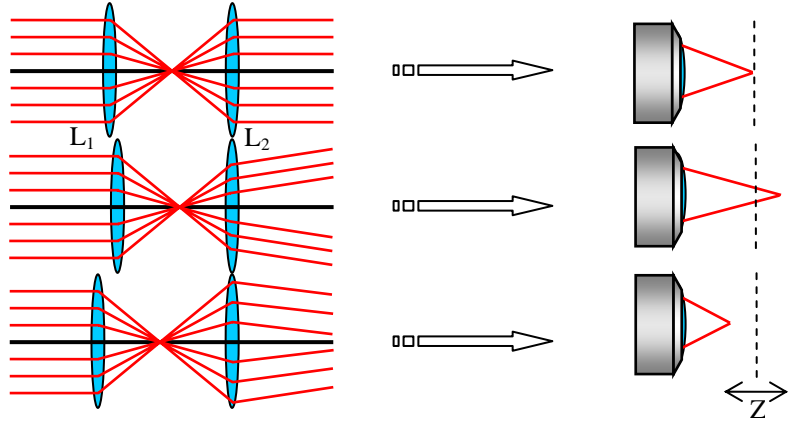

Figure 15. Illustration of how the divergence of the light is changed by a movement of the first lens, $L_1$, in an afocal optical system. This method provides adjustment of the position of the trap in the axial direction as shown at the right side.

Lateral steering of the trap is achieved by the optical system (lens $L_3$, $L_T$ and objective) together with a piezo controlled mirror[3] positioned at the image plane of the entrance pupil of the objective, as is schematically illustrated in Fig. 16. Since the entrance pupil of the objective and the mirror are in conjugate image planes, a pivot of the mirror does not change the amount of laser light entering the objective. Thus, when the mirror is tilted the trap strength is not changed. A relation between the pivot angle, $\theta$, of the GMM mirror to a lateral movement, $r$, in the specimen plane, can be expressed as [56],

$$r = -2 f_{obj} \frac{f_3}{f_T} \theta . \qquad (12)$$

A deflection of the GMM of 1 mrad will then move the trap in the lateral direction in the sample a distance of 3 μm. The resolution of the piezo mirror ~μrad, thereby allows for nm displacements in the lateral direction. This type of displacement is used during the initial part of the calibration procedure as mention below. It is also possible to construct a force feedback system, i.e., a system that can apply a constant force, by tilting the GMM mirror. A force feedback system is very useful for measuring step responses [57].

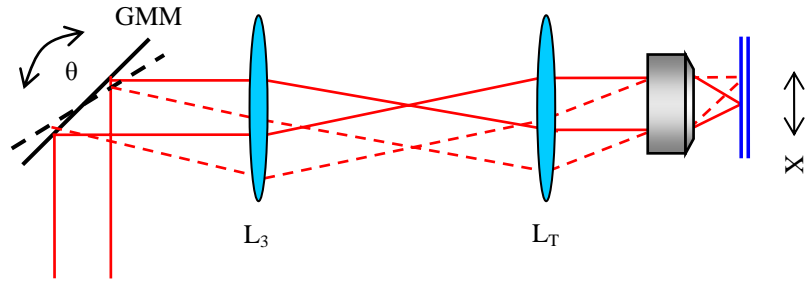

Figure 16. A setup of an optical system that allows for a controlled movement of the trap in the lateral direction of the sample plane by moving the gimbal mounted mirror, GMM that is placed at an image plane of the entrance aperture of the objective. This method prevents possible intensity losses in the objective during lateral beam steering.

To allow larger movements ($\pm$ 20 µm) of the trap in the lateral direction in the sample, the piezo controlled mirror was mounted in a step motor controlled gimbal holder[3]. This particular solution is useful during initial alignment of the trap.

In conclusion, this type of optical system provides good movability of the trap in the specimen plane and adjustments in 3-D without any loss of power by a pivoting of the laser beam around the entrance pupil of the microscope objective [56]. This recipe also reduces the influence of vibrations of the mirrors and lenses on the stability of the trap.

### 3.2.3. The influence of spherical abberation

The main advantage of using oil immersion instead of water immersion objectives is the fact that it is possible to have a higher N.A., thus the high converging rays creates a stronger axial trap for the same amount of output power. However, the index of refraction mismatch between the glass and water interfaces create spherical aberration [58-60]. This aberration distorts the focus, i.e., the focus is not a diffraction-limited volume in space, instead it is rather elongated. Since the trap strength will fluctuate with the shape of the focus, a trapped particle is sensitive to changes in the height above the cover slide,. A ray tracing simulation with an optical design program is presented in Fig. 17, which shows the influence of the index of refraction mismatch. The simulation assumes high converging rays that propagate through a cover slide into a water solution. One should keep in mind that this raytracing is build upon a geometrical optics approximation and does not take into account diffraction effects. However, it gives an illustrative picture of spherical aberrations present in optical trapping experiments.

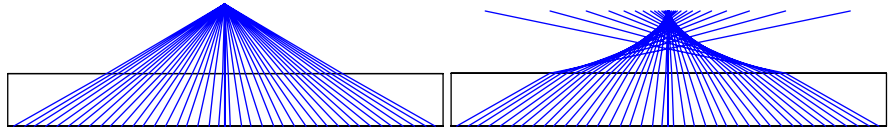

Figure 17. A ray tracing simulation that shows the spherical aberration that comes from the index of refraction mismatch between the oil/glass-water interfaces, left illustration.

The simulation above shows the importance of keeping a constant height during lateral force measurements. First, as mentioned above, spherical aberration creates a non distinct focal volume. This implies that a drift in height will severely affect the trap efficiency. Second, the wall effects of the trapped object and the cover slide, the corrected Stokes drag force, is height dependent. This implies that a drift in height changes the shear force. Therefore, our biological model system, where a large bead is immobilized to the cover slide (illustrated in Fig. 7), allows for measurements performed at the constant height. However, if the horizontal translation is not parallel to the cover slide the height and thereby the trap constant will change. We have therefore implemented a possibility to adjust the tilt of the sample so a translation in the horizontal direction is performed at a constant depth.

## 3.3. The bead position detection system

### 3.3.1. Instrumentation

As mentioned above, the detection method used in this work is built on a separate laser used for probing the position of the bead in the trap. The probe laser was a fiber-coupled[3] linearly polarized HeNe laser[3]. A fiber-focuser[3] was mounted at the end of the fiber and positioned on an x-y-z-micrometer stage[3] for accurate alignment, shown in Fig. 18. The light was directed towards a Gimbal mounted mirror[3] positioned at a 45° angle relative to the fiber-focuser. The use of a micrometer stage in combination with a Gimbal mounted mirror made it possible to align the focus of the light in a very precise manner in the x, y, z, θ, and φ directions. The light was introduced into the microscope through a side port and directed towards the objective by the use of a polarizing beam splitter cube (PBSC)[3]. The focus of the light after the fiber-focuser was positioned in close proximity to the image plane of the trapped bead in such a way that the beam focused slightly below the trapped bead in the specimen plane. The probe beam was then adjusted to provide an as large linear detector-*vs.*-bead-position region as possible. The power of the probe light was a few μW in the sample plane, thus the probe light is neither affecting the trapping efficiency nor heating the trapping volume.

The advantages of using a fiber-coupled laser are mainly that the spatial mode of the output beam is clean (it provides an excellent Gaussian beam profile), which improves both the focusing of the beam in the specimen plane and the imaging onto the detector. Furthermore, it reduces the number of mirrors and lenses that needs to be used in the system. Since each mirror and lens can introduce noise by acting as antennas for vibrations, the use of a fiber-coupled laser reduces the susceptibility to vibrations. Moreover, its long time stability is outstanding and facilitates convenient optimization since it is easy to adjust. Typically, fine adjustment of the probe beam is completed in less than a minute. In addition, the use of a HeNe laser that operates in the visible region (632.8 nm) makes aligning much easier compared to laser light in IR. Finally, a fiber-coupled laser system is in general less expensive than the corresponding number of lenses, mirrors and mechanical holders that are needed when regular optical components are used.

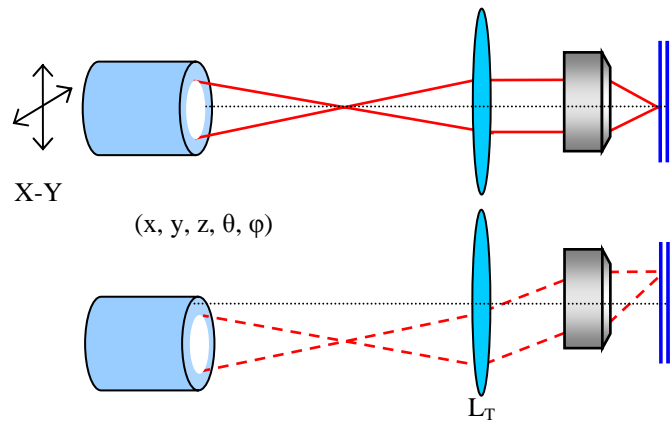

Figure 18. Moving the fiber-focuser in the lateral direction creates a movement of the trap in the sample plane. The fiber-focuser was also positioned in a way that allows for adjustment of all degrees of freedom, x, y, z, θ, and φ, with an easy and precise alignment. This type of setup replaces both methods shown in Fig. 15 and Fig. 16.

### 3.3.2. The PSD detector

As mentioned above, a trapped bead forms together with the condenser an optical system that images the transmitted and scattered light onto a detector and thereby allows for position monitoring of the object relative the trap. In this work, a PSD[3] mounted on top of the microscope in an x-y-micro meter position stage, was used as detector. The fast response, extraordinary resolution, and good linearity over the detector area make a PSD a well suited detector for the laser probe technique presented. The PSD converts incoming light to four continuous photocurrents, two for each lateral direction (X and Y), that are converted to voltages via electronic converters. It is thereby possible to measure the two lateral directions simultaneously, $X(V_x^+, V_x^-)\ Y(V_y^+, V_y^-)$, as is illustrated in Fig. 19. For light impinging at the centre of the detector, the currents from the four directions are thereby equal. Moreover, a normalized intensity independent voltage signal in each direction is created as $\Psi_x = (V_x^+ - V_x^-)/(V_x^+ + V_x^-)$. It is then possible to relate the shift of the light impinging on the detector to the deviation of the bead from its equilibrium position. Furthermore, if the detector signal is calibrated against a known displacement, as discussed below, it is possible to define a conversion factor that relates the signal $\Psi_x$ to the absolute displacement of the bead $\Delta$x.

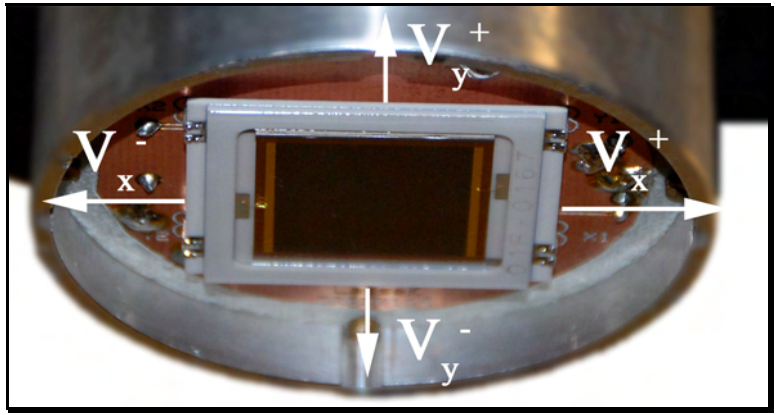

Figure 19. The PSD converts the incident probe light to a current. The current is converted to a detector signal, $\Psi_x$, which in turn, is related to the bead displacement via a signal-displacement calibration.

## 3.4. Calibration of the force detection system

The calibration of the detector signal to a bead displacement consists of two steps. The first relates the tilt of the gimbal mounted piezo mirror to a translation in the sample plane and the second relates the detector response to a movement of the bead. The first calibration needs only to be performed once if the system is well aligned. The strength of this calibration method is that it makes it possible to calibrate each bead that is used as a force transducer separately, prior to each measurement. Thus, the measured trap constant is therefore considered to be reliable.

### 3.4.1. Calibration of the lateral movement of the trap by the gimbal mounted mirror

The aim of the first calibration step is to find the relation between the tilt of the gimbal mount mirror and the lateral translation of a trapped object in the sample. To accomplish this, we use a digital imaging technique that is built upon using the high resolution translation of the piezo stage as a calibration scale. In short, the piezo stage is brought to a start position after which it is moved in several discrete steps to an end position, ~80 µm. During each step a high resolution image is captured. These images are digitally treated and used as a calibration image for later purpose. A digital image of the translation of a bead stuck to the cover slide is seen Fig. 20, where each discrete step is 10 µm, for a typical calibration the step size is 5 nm. It is now possible to relate a distance in the field of view with the number of pixels in the image. A bead is then trapped at normal working conditions and the trap is moved in the lateral direction by tilting the piezo mirror $\Delta\theta$, in discrete steps. At each step an image is captured and digitally treated. These two images make it possible to determine an angular-to-displacement factor, $\xi = \Delta x / \Delta\theta$, that can be used to find the linear response of a trapped object.

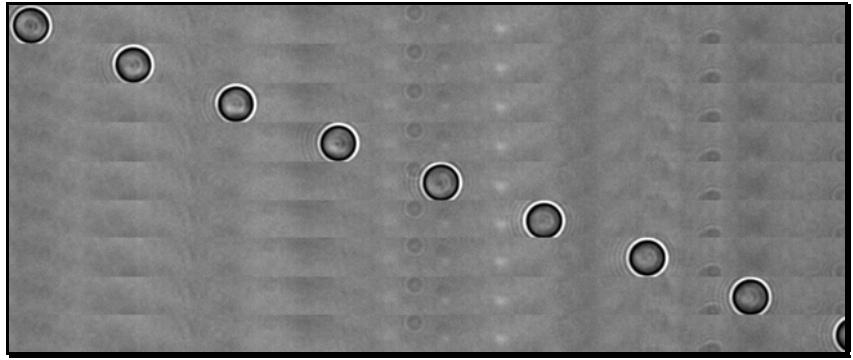

Figure 20. Digital image of the translation response of a bead stuck to the cover slide. The piezo stage was moved with discrete steps 10 nm. The image is later used as a calibration image for determination of the tilt-*vs.*-lateral-trap translation of the GMM piezo mirror.

### 3.4.2. Calibration of PSD detector response

The second calibration step relates the detector response in voltage to a movement of the bead in units of displacement. A typical probe response of a 3.0 µm bead illuminated with a probe laser is shown in Fig. 21. The measurements were performed by moving a single bead with small discrete steps by tilting the GMM mirror while keeping the probe laser stationary. The signal was recorded and averaged at each position and fitted with a polynomial (solid line) for all data points and a linear polynomial for the linear region. In a typical force calibration, we displace the bead less then half the bead radius with several discrete steps and fit a linear polynomial to those data points. The data in Fig. 21 shows that a spherical particle only has a limited linear signal range for displacements typically up to half of the radius of the trapped bead. It is, however, possible to use a longer range if a non-linear polynomial is fitted to the data, as shown with the red solid fit. However, this type of fit should be used with care since it is a possible that the optical trap does not show a linear behaviour at the wings.

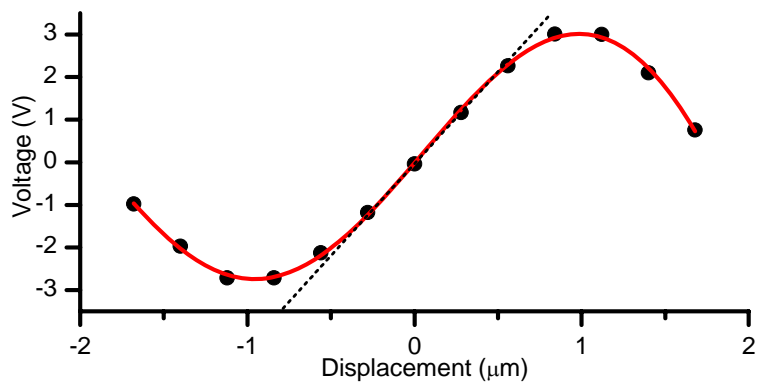

Figure 21. The data curve shows a single probe response of a 3 µm bead (diameter) in the lateral direction. The middle part of the curve represents the linear region of interest (dotted line), i.e., the part of the displacement that is used in force measurements to make sure that the force increases linearly. However, the non-linear fit (red solid line) could be used under certain conditions.

## 3.5. System control and steering

Experiments that involve nm/s-motion with nm-displacement and pN-force resolution require sophisticated hardware and software. The laboratory equipment used in this setup were carefully chosen, placed and used to minimize the influence of the operator. All types of translation and measurement were controlled via a computer interface program written in LabView. The program is written in a way that it gives the operator a possibility to align the lasers for optimized signal-to-noise ratio and to choose between numerous measurements procedures. For a screenshot of the program, see [61].

### 3.5.1. Manoeuvring of the sample

An additional closed loop nano-positioning x-y-piezo stage[3] (illustrated in panel A of Fig. 22) was mounted on top of the existing microscope step-motor stage[3] to improve the positioning and steering of the sample. It provides translation motion at a few nm/s with sub nm positioning resolution. The piezo is controlled via a computer interface for accurate movement, and the absolute position of the stage is sampled for computing the exact translation distance. A custom made mount is firmly attached to the aperture of the piezo that fixates the cover slide holder, which is made of dural aluminum. The cover slide holder and the position of the sample are illustrated in panel B of Fig. 22. A sample is placed on a 0.17 mm thick cover slide that is clamped with clips, not shown in the illustration, to prevent settling during a measurement. Finally, a top slide is carefully placed as a lid and firmly secured to seal the sample volume, ~ 25µl. A detailed protocol for sample preparation is given in Chapter 6.1.1.

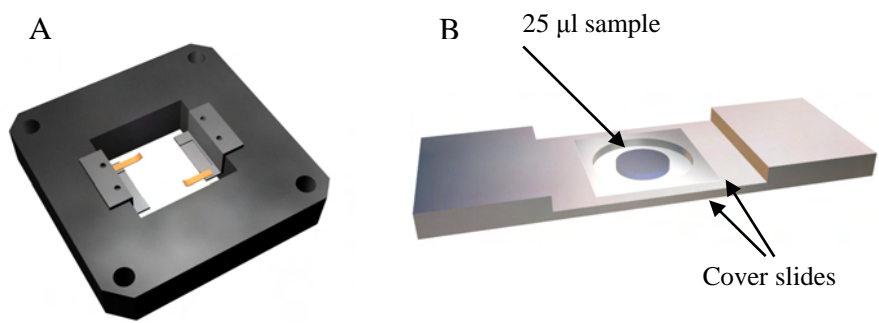

Figure 22. Both figures are part of the sample holder system. Panel A shows the modified x-y-piezo stage that holds the sample chamber. Panel B illustrates the cover slide holder that consists of a thin aluminum frame with a hole in the middle. A cover slide is secured with two clips, not shown in the figure, to prevent settling during a measurement and an additional top slide seals the sample.

### 3.5.2. Controllers and signal flow

The controllers for the piezo-x-y stage and the piezo-mirror, and the filter SR640 are steered via the GPIB bus, whereas the microscope stage motor controller[3] is connected via the RS-232 bus. The signals from the PSD detector $V_x^+, V_x^-, V_y^+,$ and $V_y^-$ are fed into separate programmable filters[3]. Each set of signals, e.g., $(V_x^+, V_x^-)$, are separately amplified and filtered. To provide an intensity normalized signal the filters are run in differential mode, i.e., the output is $(V_x^+ - V_x^-)$. Moreover, the signals are amplified since the difference signal is very small at an equilibrium position, i.e., when the light is centred on the detector surface. The detector signals are coupled to a 16 bit (1 Msample/s) A/D card[3]. The acquisition is controlled with a LabView program. Furthermore, the program is written in such a way that, depending on the operational mode; i.e., probe response, calibration or type of measurement, it automatically sets a suitable filtering of the SR640 to avoid aliasing effects. The position of the piezo stage is sampled with sub nm accuracy. Finally, a heat sensor that is connected to the sample chamber is read before each calibration procedure in order to get as accurate value of the local temperature as possible. A flow chart of the input and output signals is illustrated in Fig. 23, respectively.

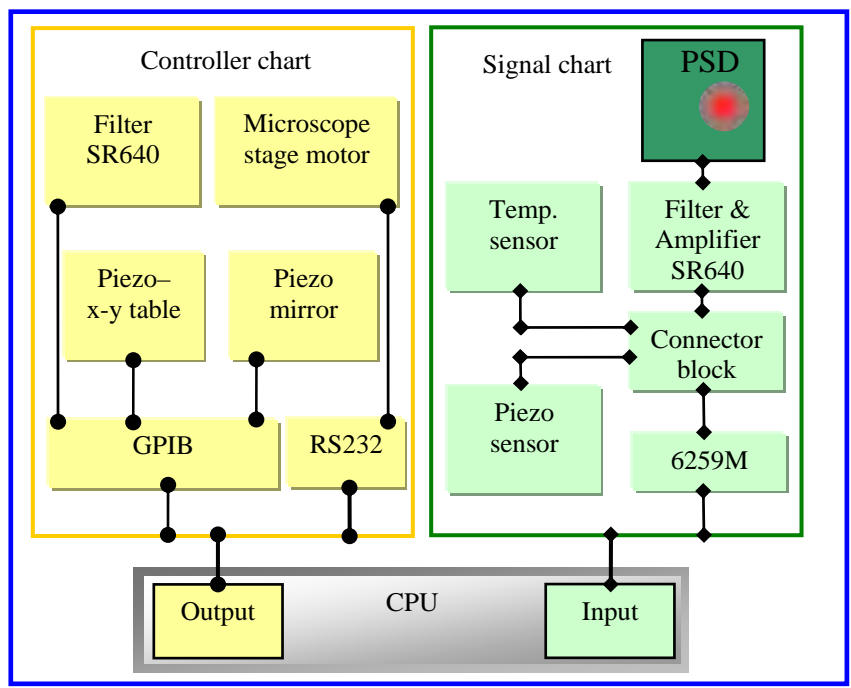

Figure 23. The left side of the schematic flow chart shows the controllers that are connected and steered by the computer. The right side shows the signal chart, i.e., the type of signals that are sampled and monitored.

## 3.6. Characterization of the instrumentation

The experimental environment (room condition) is very important for the stability of an optical tweezers. As is shown in the photo in Fig. 24, the room was constructed with a noise reducing ceiling, a separate air vent system, sliding doors, and a possibility to place controllers, computers, etc. outside of the room. This type of environment reduces air circulation, minimizes dust particles, and provides good control of the temperature. However, analysis of the stability of the FMOT system, lasers, temperature, and mechanical components is checked on a regular basis to make sure that the setup is not drifting in stability. Such analyses are e.g., power linearity, power spectrum analysis, converting factors, and degree of entrance aperture overfilling.

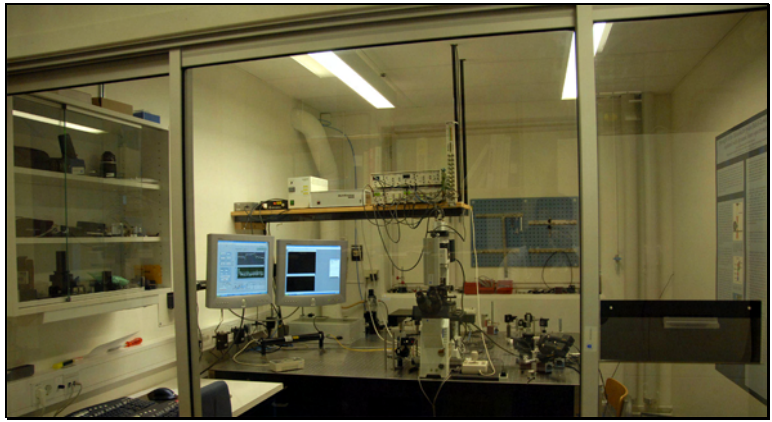

Figure 24. A photo of the experimental setup that is placed inside a room that is designed to minimize the air fluctuations, temperature gradient, and noise.

To assess the linearity of the trapping force, i.e., to measure the trapping constant for different laser powers, measurements were made, which according to Eqs (1) and (5) should show a gradient force proportional to the intensity of the light. A series of measurement, where the power was increased and the trap calibrated by the Brownian motion method, was performed and plotted in Fig. 25. As can be seen in the figure, the trap strength increases linearly with output power, which indicates that the overfilling of the entrance pupil is sufficient and that the laser beam profile is not changed drastically when the power is altered. The good linearity of the power thereby makes possible measurements where the loading rate is changed by altering the trapping constant instead of the extension speed.

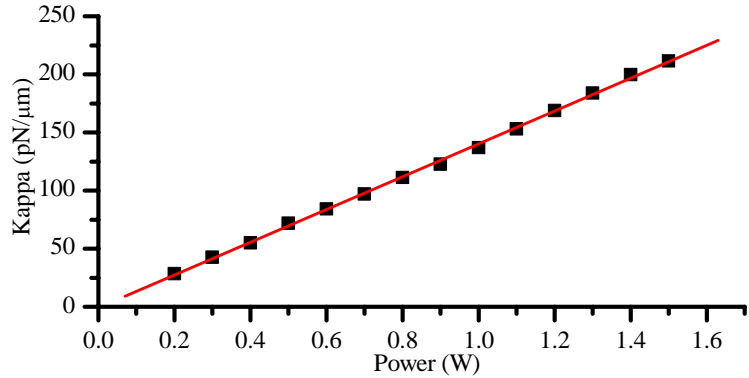

Figure 25. Trap strength-*vs.*-output power of the laser. A trapped bead was calibrated at different output powers of the laser. The linearity of the trapping strength for different laser powers shows that the system is well aligned.

An advantage of calibrating with the power spectrum method is that the condition of the trap is immediately shown, i.e., the amount of noise for each frequency is shown. Thus, it is possible to characterize mechanical low frequency noise and electronic high frequency noise. Figure 26 shows three typical power spectra for measurements at 0.5, 1.0, and 1.5 W of output laser light. A fit with Eq. (8) resulted in trap constants of 67, 142, and 223 pN/µm, respectively.

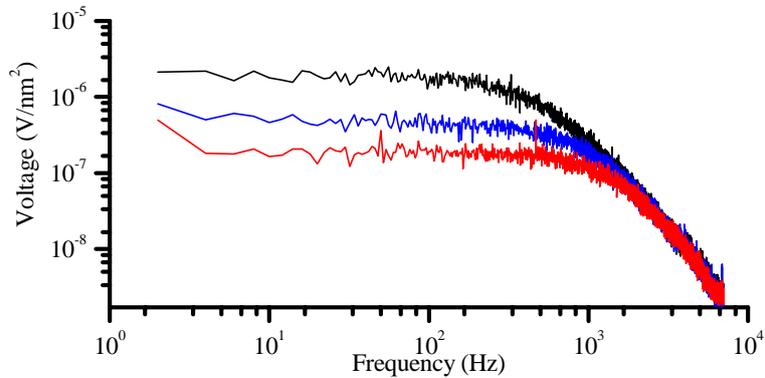

Figure 26. Three typical power spectra of a bead trapped at 0.5, 1.0 and 1.5 W of laser power. A fit with Eq. (8) to the data gave trapping constants for these particular measurements of 67, 142, 223 pN/µm, respectively. The data has been smoothed for presentation purposes.

As can be seen from the figure the low frequency noise in the trap is small, i.e., the noise below 10 Hz is almost constant. The validity of the power spectrum method was also compared with Stokes drag force calibration, data not shown. Moreover, in Fig. 27 the position distribution of a trapped bead at two different laser powers, 200 and 800 mW, are shown. Such ensemble of the position distribution can be described by the Boltzmann distribution as, $P(r) = A\exp(-kr^2/kT)$ [62]. A fit with this equation to these data sets resulted with a spring constant of 36.4 and 168.3 pN/µm, respectively. The same data was also fitted by the power spectrum method which only differed within 2 %.

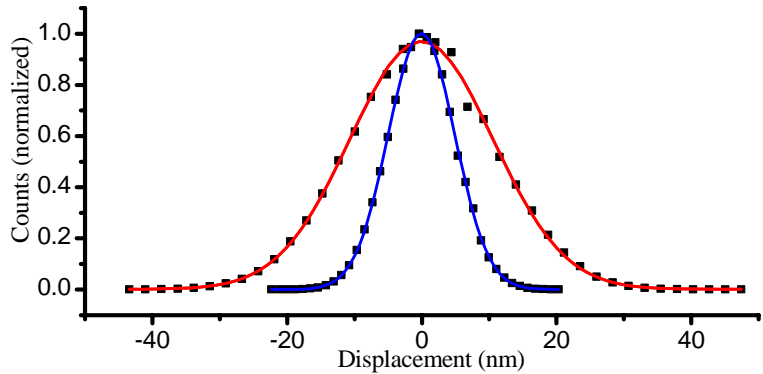

Figure 27. A normalized histogram plot of a trapped particle at 200 mW (wide red distribution) and 800 mW (narrow blue distribution) of output power. As seen from the data, the stronger trap confines the particle motion and restricts the visits at the wings. The trap strength was determined to 36.4 and 168.3 pN/µm, respectively.

Since many measurements involve continuous experiments of the order > 10 min, a lot of time has been spent to create a long time stabile tweezers system. For that purpose, the low frequency drifts, i.e., mechanical vibrations, temperature fluctuations, and lasers instabilities, were investigated and decreased. Figure 28 shows a long time measurement with a trapped bead in its equilibrium position. The slow moving fluctuations originate presumably from the mentioned factors. However, the peak to peak fluctuations of the slow moving drifts are < 0.2 pN.

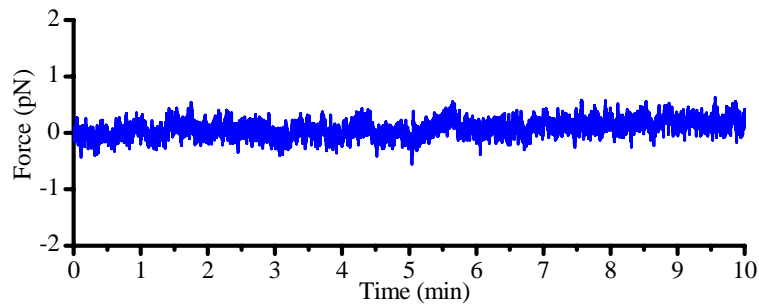

Figure 28. A 10 min force measurement of a trapped bead. The long time drifts are in general small and the fluctuations in force < 0.2 pN.

Table 1. Specification list.

| Component | Manufacturer | Series name or number |
|---|---|---|
| Optical table | TMC | Custom made optical table |
| Microscope | Olympus | IX-71 |
| Nd:YVO$_4$ laser | Spectra Physics | Millennia IR |
| IR coated dichroic mirror | Laser components | HR1064HT633 + 337/45/BBAR RW33-25.4-3UV, SWP-L-33x1 |
| Objective | Olympus | UPLFL100X/IR, NA 1.30 |
| L1 | Melles Griot | f = 60 mm, LPX 125/083 |
| L2 | Melles Griot | f = 60 mm, LPX 125/083 |
| L3 | Melles Griot | f = 150 mm, LPX 238/083 |
| Blocking filter | Omega Optical | Hot mirror, 840 |
| Piezo mirror | Physik Instrumente | Model S226 |
| Step motor | Melles Griot | Model no. 07 TSC 517 |
| Fiber-coupler | OZ Optics LTD | SMJ-A3A, 3AF-633-4/125-3-1 |
| HeNe-laser | Manteca | 1137, Uniphase |
| Fiber-focuser | OZ Optics LTD | HPUFO-2, A3A-400/700-S-17-180-10AC |
| Micrometer stage | New Focus | 9064-XYZ |
| GMM | Melles Griot | Model no. 07 MCD 515 |
| PBSC | CVI laser Corp. | 50/50 |
| PSD | Sitek | Model no. L20 SU9 |
| Piezo stage | Physik Instrumente | P517.2CL |
| Step-motor stage | Märzhäuser | |
| Preamplifier / Filter | Stanford Research system | SR640 |
| Connector block | National Instrument | BNC2090A |
| A/D-card | National Instrument | PCI 6259M |
| CCD camera | Kappa | DX-20h / DX-2h |

# 4. PROBING SINGLE MOLECULES

Biological physics is an interdisciplinary field that uses physics as a tool to describe questions and phenomena that appear in biological systems. Furthermore, the field is very broad since it includes issues from a number of more classical fields, e.g., biology, biochemistry, chemistry, physics and mathematics. The field, thus, forms a new and important research field by itself. The development of refined experimental techniques and fast computers has opened a new possible way to study, explain and verify many different biological processes. Both static and dynamic processes have been explored and the functionality of some of the central components of life, the proteins, has been investigated and is still an important field for research. The biophysical work, presented in this thesis, focuses mainly on larger attachment organelles, so called pili. However, before looking into these in detail, let us first provide a brief introduction to the basic elementary structures of biological systems.

## 4.1. Elementary structures

The fundamental and necessary building blocks of all living matter are proteins. They exist in a variety of forms; enzymes, hormones, receptors, transporters, antibodies, molecular motors, and structural proteins [14]. Moreover, they are complex architectures that are assembled with a few or a multitude of amino acids in a sequence, typically between 50 to 200 units. As is shown in Fig. 29, each amino acid consists of an amino group (-NH$_2$), a carboxyl group (-COOH), and a side chain (R), all bound to a carbon atom. There are twenty different side residues that classify the different amino acids. Polymerisation of amino acids occurs via a peptide bond, i.e., a covalent bond (dotted line) is formed between the carboxyl and amino groups. A linear chain of such amino acids is called a peptide. A protein is composed of numerous linked peptides (polypeptide), which has a specific fold mainly govern by its side chains. Interestingly, it is the structure of the protein that reflects the function, which in turn is connected with the rigidity of their three-dimensional architecture [14].

The final protein is typically composed from a few low ordered structures. The *primary structure* is the linear conformation of the different amino acids into a polypeptide chain. The *secondary structure* is polypeptide regions that are locally folded into regular stable structures known as α helices or β sheets. These secondary structures can create *tertiary structures* that form the complete picture of the three-dimensional protein. Finally, the *quaternary structure* is the result of multiple folded polypeptide chains into a multi-subunit complex.

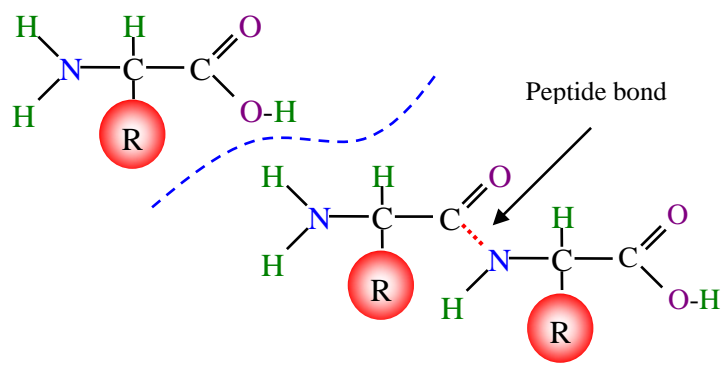

Figure 29. The upper left shows a schematic illustration of a single amino acid whereas the lower (below the dashed line) shows the formation of a peptide via a peptide bond (dotted line). The filled circle (R) represents the side chain of the amino acid.

## 4.2. Single molecule measurements

Traditionally, molecules and chemical compounds have been investigated via bulk experiments, e.g., calorimetry measurements, that give time and population averages of individual molecules close to equilibrium [15, 63]. Such measurements provide the most probable state of the molecules but fail to see the intrinsic behavior of an individual [64, 65]. However, the development of new probing techniques, e.g., optical tweezers, AFM and bio-membrane probes, has made it possible to perform single molecule experiments [66, 67]. Such single molecule measurement techniques allow investigation of; small structural changes, intermediate transition states, and details of an energy landscape without exploring an ensemble [68]. Additionally, the 3D-architecture of proteins, cell adhesion strength, forces in chemical reactions, behavior of molecular motors, biochemical kinetics etc, are also targets of interest that can be probed with success [15]. In conclusion, single molecule techniques are able to explore, probe and measure interactions at a resolution that traditionally has not been accessible.

Even though single molecule force spectroscopy can reveal information that traditionally has not been accessible, there are two reaction regimes that must be considered. These are reactions that proceed via equilibrium and non-equilibrium processes. In quasi-equilibrium reactions, it is possible to extract the free enthalpy from a force-extension curve [69]. Such processes are reversible and there is therefore no hysteresis between the extension / contraction curves. The system can thereby access all configurational states on the time scale of the experiment. However, if the experimental time scale is faster then the inverse of the transition rates, the equilibrium condition is violated and the measurement is performed under non-equilibrium conditions. In such case, the applied force causes a hysteresis and the extension / contraction curves will not coincide. However, force dependent reactions open up an opportunity to explore and extract information about molecular interactions by allowing for measurements under non-equilibrium conditions.

## 4.3. Kinetic description of bond transitions under applied force

In the microscopic world of molecules, it has for a long time been known that molecules with an excess of energy above the average energy participate in a reaction [70]. The dissociation of a molecule complex (e.g., a receptor-ligand bond) follows a pathway along a reaction coordinate, where the rate of reaction is controlled by the surrounding environment and a transition barrier. For a two state system, a bound state A and unbound state B are separated by a single energy barrier referred to the transition barrier. These states could represent a receptor-ligand complex where an energy landscape, defined by the energy height $\Delta V_{AT}$ and the potential width $x_{AT}$ of the transition barrier, controls the dissociation time of the complex. An example of such a two state system is illustrated in Fig. 30.

The natural dissociation rate can be expressed in terms of the height of the transition barrier and the thermal energy $kT$ as,

$$k_{AB}^{th} = \nu e^{-\Delta V_{AT}/kT} , \qquad (13)$$

where $\nu$ is the attempt rate. In addition, the association rate of the reaction can then be defined in a similar way,

$$k_{BA}^{th} = \nu e^{-\Delta V_{TB}/kT} , \qquad (14)$$

where $\Delta V_{TB}$ is the energy difference between the unbound state and the barrier.

In 1978, Bell introduced a simple model of force induced bond failure based on a transition-state theory for receptor-mediated cellular adhesion [71]. The model assumes that the energy of the transition barrier is reduced to $\Delta V_{AT} - Fx_{AT}$ when a mechanical force acts along the reaction coordinate, as illustrated by the red dotted line in Fig. 30. The reaction coordinate is defined along the direction of the applied force. The reduced barrier increases the dissociation rate and the force dependent dissociation rate can then be expressed as,

$$k_{AB}(F) = \nu e^{-(\Delta V_{AT} - Fx_{AT})/kT} . \qquad (15)$$

Moreover, the force increases the energy barrier for association and the force dependent association rate can therefore be expressed in a similar way,

$$k_{BA}(F) = \nu e^{-(\Delta V_{TB} + F x_{TB})/kT} . \quad (16)$$

From this point, it is convenient to define Eq. (13) as the thermal rate, since this allows Eqs (15) and (16) to be written as,

$$k_{AB}(F) = k_{AB}^{th} e^{F x_{AT}/kT} , \quad (17)$$

$$k_{BA}(F) = k_{AB}^{th} e^{(\Delta V_{AB} - F x_{TB}/kT)} . \quad (18)$$

where $\Delta V_{AB} = \Delta V_{AT} - \Delta V_{TB}$. These rates change exponentially with the applied force which is the trademark of Bell's theory. Thus, by applying a mechanical force it is possible to change the rates and thereby decrease the natural lifetime. Unbinding forces can then be measured since the time-span until rupture is short.

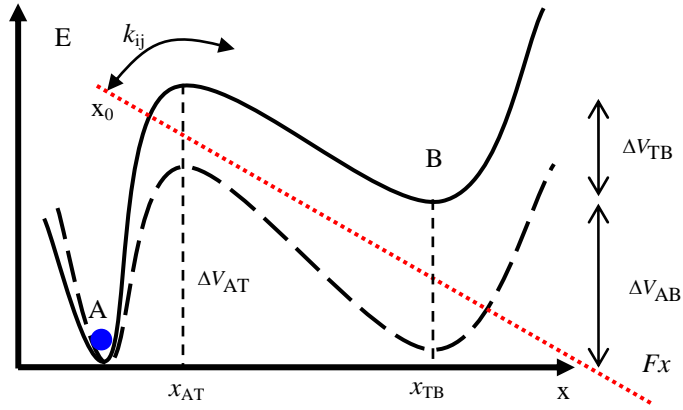

Figure 30. A schematic illustration of a two-state energy landscape where a mechanical force (red dotted line) is applied along the reaction coordinate. The mechanical force changes the intermolecular potential and the natural bond lifetime is decreased. Notice that the applied force is zero at the initial position, $x_0$.

In conclusion, a rate dependent process in the microscopic energy landscape, where the user can control the rate by mechanical force, can be explored if the force transducer is sensitive. The theory of Bell is used in this thesis to model the force dependent bond opening rates of both the layer-to-layer bond and head-to-tail bond of subunits in pili.

Evans refined Bell's works and developed a theoretical model for how to use and interpret results from experimental measurements on single bonds [31]. However, in this thesis, single bonds have not been investigated, but a detailed description of his theory is presented in the following references [28, 31].

# 5. BIOLOGICAL MODEL SYSTEM

Infection diseases are common and millions of people are each year infected by different species of pathogenic bacteria, many with deadly outcome. Infections have been intensively and successfully treated with antibiotics during the second half of the 20$^{th}$ century. However, the intensive use of antibiotics has unfortunately resulted in a widespread bacterial antibiotic resistance, resulting from bacterial mutation or an ability to acquire resistive genes from other bacteria through conjugation [72]. Therefore, treatment of infections can nowadays be a challenging and difficult problem. As a consequence, there is an urgent need for development of new strategies to defeat infections.

The initial step of an infection starts via the bacterium's ability to adhere to target cells. Bacterial adhesion can therefore be seen as a key virulence mechanism [73]. Adhesion is mediated by molecular interactions between the adhesin (a compound protein structure) on the bacterium and the receptor of the host cell. Since both bacteria and cell membranes are negatively charged they will repel each other when being in close proximity [74]. As a consequence, bacteria have developed different structures or organelles for expressing the adhesive protein distal from its surface. These surface organelles, called fimbriae or pili, are in many cases indispensable for the initial contact and resistance to various types of cleaning actions. A promising and alternative way of fighting an infection is therefore to find new drugs that can address the adhesion mechanism. However, in order to develop such drugs, a better understanding of the complicated molecular mechanisms of bacterial adhesion must be obtained.

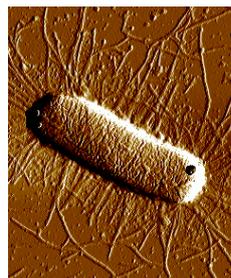

The pathogenic variants of *Escherichia coli* (*E. coli*) bacteria serve as an important model system for investigation of general bacterial adhesion to host tissue [75]. The AFM picture to the right shows an uropathogenic *E. coli* (UPEC) bacterium expressing a multitude of pili. UPEC are usually equipped with pili and classified as the major pathogen >80 % in urinary tract infections (UTI). Millions of people are each year infected with UTI and recurrent infections are common [76]. The medical cost is huge, and antibiotics are most often used as the antibacterial drug. Therefore, a thorough understanding of their intrinsic adhesion organelles will help scientist in the search for more effective drugs. It is possible that the use of FMOT as a tool to assess various properties of adhesion organelles on an individual basis can speed up the development of such drugs. This thesis contributes to this field of science by presenting the first characterization of the biomechanical properties of some typical members of the pili family under natural conditions.

## 5.1. Pili, the key to firm adhesion

As mentioned above, pili structures mediate the initial contact with host cells. Thus, they provide the important receptor-ligand interaction and are necessary for firm adhesion. Moreover, evolution has developed these organelles to sustain natural host defenses, e.g., urine flow [77]. This flow removes effectively unbound bacteria and those with weak receptor interactions. However, it is a general consensus that the large flexibility of pili is a crucial property of bacteria to sustain the rinsing action from urine flow. The flexibility presumably allows distribution of the urine shear forces applied to the bacterium over a large number of organelles [78, 79].

P and type 1 pili are well characterized and investigated members of the pili family. They are composed of numerous subunits connected via a donor strand complementation, i.e., every subunit donates its amino terminal extension to complete the fold of its neighbor through a non-covalent bond [80, 81]. P pili are predominantly expressed by isolates associated with severe infections in the kidney, more known as pyelonephritis [82, 83], whereas type 1 is expressed by 80 % of all wild-type *E. coli* and responsible for establishment of cystitis[4] [77].

These structures are 6-7 nm thick and typically around a fraction to a few µm long [84, 85]. Moreover, electron microscopy images have shown that P pili are composite structures made from many subunits. The main part of this extracellular organelle, the rod, is composed of repeated PapA units arranged in a right handed helix [86], with a pitch of 2.49 nm and 3.28 subunits per turn [87]. At the end of the rod a thin thread like fiber is attached (tip fibrillum) via an adaptor PapK. The fibrillum is made up of a short homopolymer of PapE, which has a second adaptor, PapF, that joints the adhesin PapG. PapG binds to the glycolipid receptors expressed by erythrocytes and uroepithelial cells [88, 89]. Finally, PapH anchors the extracellular organelle to the membrane which terminates the assembly process. An illustration of the structure of a P pilus is given in Fig. 31.

Comparable images of type 1 pili show structural similarities with P pili, i.e., the rod is composed of FimA subunits arrange in a helix-like way with a pitch of 2.31 nm and 3.125 subunits per turn. The fimbrial structure contains the two adaptor proteins FimF and FimG and the adhesin FimH. The FimH adhesin has a carbohydrate-binding pocket that binds to mannose-containing glycoprotein receptors expressed by host cells [81].

---

[4] Inflammation in the bladder.

The properties of the adhesins, PapG and FimH, have not been investigated in detail in this thesis. However, the major differences between the two are, except that they bind to different receptors, first the location of the binding pocket. The PapG binding pocket is located at the side of the adhesin, which therefore requires a fibrillum tip that is long and flexible [90]. In contrast, FimH has an acidic pocket that is found on the tip of the adhesin [91]. Moreover, it has been shown that the type 1 fibrillum is short and stubby, which could possibly be correlated with the distal binding pocket. Finally, it has been shown that the FimH adhesin is of a catch bond type, i.e., the structure of the protein is changed during load, thus increasing the bond strength [92, 93]. As a consequence, urine flow and bladder contraction enhance FimH binding [75].

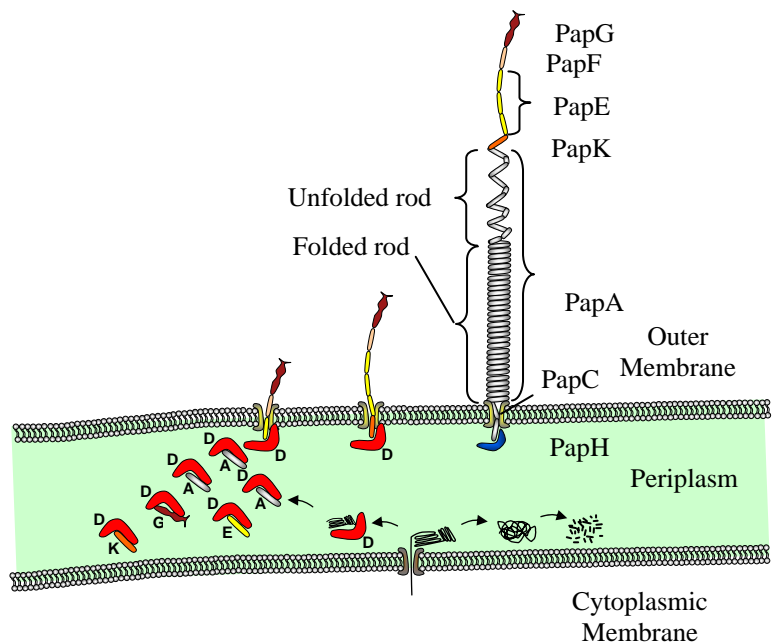

Figure 31. A schematic illustration of the biogenesis of a P pilus. These pili are assembled via the chaperone (PapD) / usher (PapC) pathway. The main part of the rod is composed of PapA subunits, whereas the tip fibrillum consists of PapK, PapE, PapF, and the adhesin, PapG. Finally, PapH anchors the pilus to the membrane and terminates the assembly process. This illustration is a reprint from ref [94].

The biogenesis of P and type 1 pili is a complicated process predominantly steered via the chaperone/usher pathway, as is shown in Fig. 31 [95]. This pathway includes a transport of a subunit from the inner to the outer membrane by the chaperone (PapD for P pili), a protein complex built from two immunoglobulin-like beta barrels. These structures create a protective cleft that stabilizes the subunits during transportation. In addition, the chaperone is not only a protein that acts as a carrier; it prevents premature aggregation of the subunits in the periplasm by donating its edge beta strand to complete the fold of the subunit [96]. The final assembly process takes place at the outer membrane via the usher (PapC, for P pili) a 2-3 nm channel that joins the individual subunits to a helix-like fiber, the pilus [97]. Since the channel is only a fraction of a helix-like fiber in width, it is apparent that individual units are translocated through the channel and assembled to a pilus at the outside of the membrane.

# 6. CHARACTERIZATION OF PILI WITH FMOT

## 6.1. Preparation and measurement procedures

### 6.1.1. Sample preparation

This section describes in short the sample preparation and measurement procedures. A procedure was designed that provided reproducible experimental conditions. Before each experiment the sample under investigation, denoted the biological model system, is prepared between two 170 µm thick cover slides, as shown in Fig. 22 Panel B. The biological model system consists of bacteria, force transducer beads (3 µm in diameter), and mounting beads (9.7 µm in diameter). The mounting beads are immobilized to the bottom cover slide, and serve as a pedestal to provide a constant measuring depth as discussed in chapter 3.2.3. To immobilize the beads to the cover slide they are functionalized by the following procedure. First, the cover slides are cleaned with filtered high pressure air to remove dust particles. A solution with filtered MilliQ $H_2O$ and diluted 9.7 µm beads (Dynamo Scientific Corp.) are dropped onto the slide s (25 µl / slide), which then are placed in an oven for 60 minutes at 55° C. This procedure distributes the beads evenly and immobilizes them to the surface.

The next procedure is to prepare the mounting beads with biological "glue". The functionality of the glue is very important since the glue fixates the bacteria firmly to the mounting beads. We have developed a method that provides a faster, more reliable, and simpler sample preparation than the method presented in [47]. In short, a solution of $H_2O$ (250 µl per slide) and Poly-L-Lysine (Sigma Aldrich) is prepared (2 µg Poly-L-Lysine / slide), mixed, and dropped onto each cover slide. The slides are then placed in an incubator for 60 minutes at 37° C. During this time the Poly-L-Lysine settles down and a positive charged surface is created. The superfluous solution is removed after the incubation by the use of filtered MilliQ $H_2O$, which is gently rinsed over slides. Finally, the slides are dried with argon.

The sample under investigation, consisting of a functionalized cover slide mounted to a dural aluminum sample holder, diluted bacteria, and force transducer beads, is prepared in the following order. First a drop (25 µl) of filtered phosphate buffer solution (PBS) is dropped onto the cover slide area containing the immobilized beads. A small amount of bacteria, scratched from an agar plate, is diluted with PBS in a 1 ml eppendorf tube. From that tube, 1 µl is withdrawn and carefully injected into the PBS solution placed on the cover slide. The 3 µm force transducer beads (Dynamo Scientific

Corp.), diluted in filtered $H_2O$, is carefully injected into the sample. The sample volume is secured with a top cover slide.

The sample holder is placed on the piezo stage bracket in the microscope. The condenser and field aperture are aligned with the standard procedure to set Köhler illumination, i.e., optimal illumination of a sample in bright field microscopy. Thereafter, a free-floating bacterium is trapped by the optical tweezers with reduced power to prevent cell damage. The bacterium is fixated at the middle part of a mounting bead. Subsequently, a small bead is trapped with the power used during the measurements, and brought in proximity to and aligned with the bacterium. The transducer bead is calibrated with the Brownian motion method, described in chapter 3. Such calibration procedure takes ~45 s.

### 6.1.2. Force-extension measurement procedures

To measure interactions at a single molecule level the operator must be careful during the initial phase of contact. The limited resolution of the microscope, typically ~500 nm, prevents the user from resolving individual pili, which are of the order of a few nm. Therefore, the bead is brought in proximity of the bacteria and swept in a sick-sack motion until the bead shows a contact response. After an initial contact, the data acquisition is started and the piezo stage set in motion to separate the bacterium from the small bead. The position-sensitive detector signal (which provides information about the force) and the output signal from the piezo driver (which provides information of the distance between the bacterium and the trap) are sampled. The detector signal is sampled and filtered to fulfil the Nyquist criterion and avoid anti-aliasing effects.

### 6.1.3. Dynamic and relaxation measurement procedures

To probe the dynamic or relaxation response of a pilus, the previous force-extension procedure is first used to map the extension behavior and reduce the number of attached pili. Since the data is captured in real time, it is possible to set start (a) and stop (b) coordinates for the dynamic measurements as is illustrated in the insert figure in Fig. 32. The measurement procedure is performed as follows. The data acquisition is started at (1). A computer program translates the piezo stage over the preset distance (2), i.e., between the two coordinates (a) and (b). The responding force rises if the motion is above the steady-state speed. At the stop coordinate (b) the piezo stage is stopped and the pilus relaxes to its steady-state level (3). The stage is reversed to the start coordinate (a) at a steady-state speed to allow the pilus to refold. Thereafter the same extension procedure, i.e., involving elongation of the same part of the pilus is carried out with an increase extension speed.

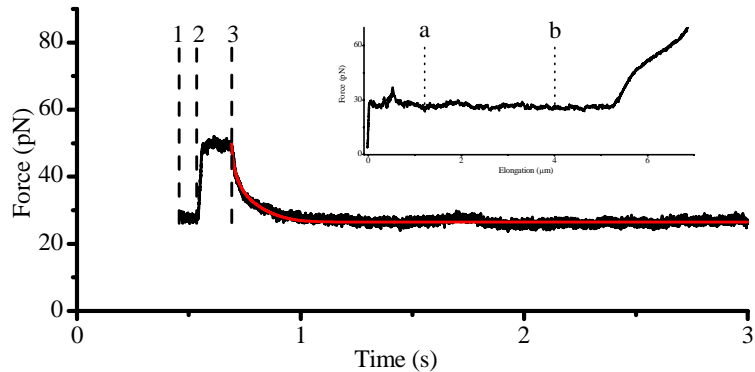

Figure 32. A response curve from a pilus during dynamic and relaxation measurements. (1) the data acquisition is started still with the piezo stage stationary, (2) the piezo stage is translated between the start and stop coordinate (a) and (b) whereas the pili responds to the increase extension speed with a dynamic response, (3) the translation of the piezo stage is stopped and the pilus relaxes down to the steady-state force. The solid fit corresponds to a numerical simulation of the complete rate equation that describes the relaxation process from a dynamic response.

## 6.2. Characterization of P pili

### 6.2.1. Introduction

As mention in chapter 6, pili are important virulence factors. Therefore, it is important to characterize and understand the intrinsic function of these structures. Until recently, the pili structure, in particular P pili, had mostly been investigated by the use of scanning transmission electron microscopy [84]. Although these static images provide information of the architecture, they do not provide information of the dynamic function in *in vivo* conditions. As a matter of fact, such "static" images can easily be misinterpreted, as is discussed below.

A three-dimensional reconstruction of the P pili architecture was presented by Bullitt et al. [78]. They showed that the main rod is formed by a tight winding of sequential PapA subunits into a coiled rod, analogous to a telephone cord. The rod is stabilized via layer-to-layer bonds between consecutive layers giving rise to a ~7 nm thick quaternary structure. Moreover, the authors proposed, from the images, that mechanical shear could unwind the rod to ~five times its original length without depolymerising the subunits. These unraveled 2.5 nm thick fibers are held together via head-to-tail bonds (complementary strands), thus forming thin curved threads. The unwinding was also believed to be plastic, i.e., once unwinded a pilus would not go back to its original shape. This assumption was based upon their static images.

Already at that time, the unfolding property was hypothesized to be a very important ability for the bacteria to withstand fluid shear forces. Simulations of how the elasticity (unfolding property) influences the bacterial adhesin were recently performed by Miller et al. [98]. Their simulations showed that the unfolding of the rod prolongs the lifetime of the bond. Thus it is possible that the rod has codeveloped with the adhesin. In addition, since PapA is the major component of the P pilus, most of the mechanical properties should be governed by the unfolding properties of the PapA rod. It is therefore justified to scrutinize the biomechanical properties of the P pilus rod.

Recently it was shown that P pili could be unfolded by a mechanical force *in situ*. Moreover, a mathematical model for the force-extension behavior was developed [99]. In contrast to the assumed plastic behavior of pili, it was shown that the unfolding of the helix-like structure is fully reversible [79]. A kinetic description of a P pilus, under steady-state and non-equilibrium stress, has thereafter been developed and is presented in **Paper I, II,** and **VI.** It is shown that the kinetic approach reproduces the behavior of pili very accurate.

Currently, the group is also investigating the adhesin, PapG, and hopefully information about its properties will then add to the understanding of the correlation between the rod and the adhesion.

### 6.2.2. A model of a helix-like polymer under steady-state extension - Paper I

A typical P pilus force-extension response assessed by force measuring optical tweezers is shown in Fig. 33. For clarity, the response is divided into different regions depending on the mode of stretching. Region I is the elastic stretching of the pilus, II the sequential unfolding, and III the phase transition of the linearized rod. All regions can be identified in the insert if Fig. 33. Based upon the extension response and the architecture of the pilus, a model of the force-extension response of a pilus was developed. It was found that the model could reproduce the experimental data very well. The concept of the model is presented in detail in **Paper I**.

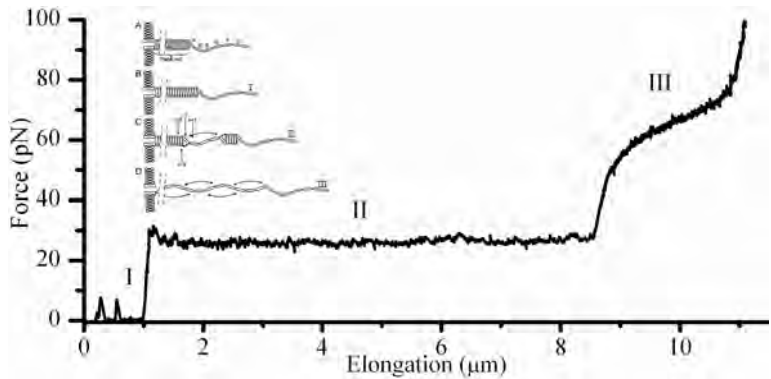

Figure 33. The plot shows a typical extension response of a P pilus under steady-state. The data is divided intro three regions, where each region is illustrated in the inline figure. The unfolding force of the quaternary structure is 28 pN ($10^{-12}$ N) and the fluctuations around the mean unfolding force ~0.5 pN. The data is sampled at 200 Hz.

Certain approximations were made to simplify the model. The first approximation assumes that a pilus is straight in its native configuration, i.e., all types of undulation and bending is neglected. The validity of this approximation is supported both by electron micrographs and through our experimental measurements [3, 86]. The measurements, performed by our group, showed that the initial elastic extension of a pilus from its native form, region I, has a straight incline. Such response indicates a long persistence length of the quaternary structure, i.e., a persistence length equal to its contour length. Secondly, a quaternary structure implies directly that the geometry of the architecture is three dimensional. Despite this, to reduce the complexity of the architecture, we neglected any influences of the three dimensional bond torsions. Instead all bond torsions are compiled to elastic constants, where some presumably could involve both torsion and elongation interactions. Finally, since the rod is composed of numerous subunits, that have both head-to-tail and layer-to-layer interactions, we created two discrete models, one valid in the regions I and II, the other in region III.

The sticky-chain model concept is a model based upon Hooks law for extension of tandem elements and a kinetic description for bond opening and closure. The model is build upon that a large number of homogenous monomers, $N_{tot}$, are attached via head-to-tail interactions ($n+1$) to a long backbone, as illustrated in Fig. 34 and shown in the insert image in Fig. 33. Moreover, each monomer is also connected to its ($n+3$)th neighbor via a so called layer-to-layer bond in the native state. Thus, the first bond to open in such structure, under stress, is the layer bond.

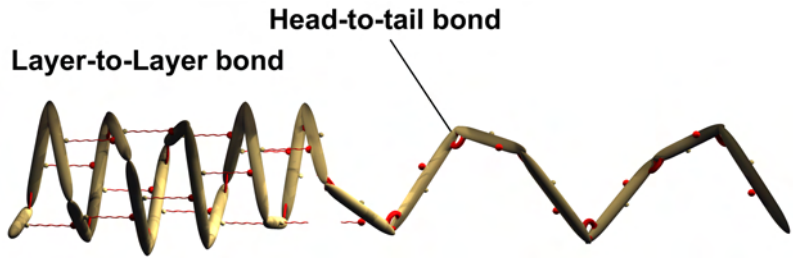

Figure 34. An illustration of how the bonds in a pilus are modeled. The red dots indicate the binding point between layer-to-layer bonds, whereas the arcs represent the head-to-tail bonds. This illustration is a reprint from ref [94].

Without strain the pilus will be in its native configuration where all monomers are in their lowest state, i.e., all monomers interact via the layer-to-layer bonds. As the pilus is elongated, the stress in the system will increase, whereby the layer-to-layer bonds between consecutive layers in the helical structure will elongate elastically. The elastic constant per unit was assessed to 61 pN/nm. As the force increases in the structure the load on each bond increases and the potential is altered, according to Eq. (15). Thus, at some point the helical structure starts to rupture.

The mode of rupture is strongly governed by the quaternary structure of the pilus. First, the helical structure of the PapA rod is mediated by several (~3) layer-to-layer bonds, whereas the outermost subunit in the helical structure is attached with only one bond. An external force, $F_{\text{ext}}$, applied to the pilus is therefore distributed among several bonds in the interior of the rod, whereas the outermost bond experience a larger force. Secondly, the opening of a layer in the interior of the pilus requires a simultaneous opening of at least three layer-to-layer bonds, which is highly unlikely. Consequently, it is only the outermost bond that will open. Therefore, the extension response for region I and II are straight responses, where the former, shows a linearly increasing response, whereas the latter gives rise to a constant force.

The good force resolution of the optical tweezers system allows for a detailed study of the linearized pili. As seen in Fig 33. data from region III indicates a pseudo-elastic behavior. Such response indicates a possible phase transition (a bond flip or an intermolecular change) of an individual subunit or the complementary strand. Moreover, since the rod is completely unfolded each monomer has the same probability to make a transition at a specific force and the pseudo-elastic response of region III is therefore attributed to the random opening of bonds. A numerical solution of the two rate equations, that defines the number of bonds in each state for a specific force, given in **Paper I,** are plotted in Fig. 35.

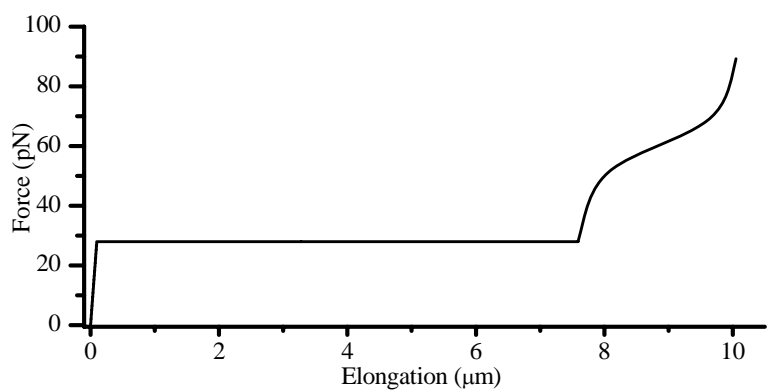

Figure 35. A numerical solving of the equations presented in Paper I under steady-state conditions. A comparison with Fig. 33, which shows experimental data, illustrates that the model describes the behavior of a P pili to an external force in the regions I, II, and III extremely well.

A comparison between Fig. 33 and the simulated model in Fig. 35, shows that the model, consisting of one collective and one individual model, presented in **Paper I,** excellently reproduces the experimental data for all three regions. The number of subunits in the pilus could be derived from region III [100]. Moreover, since the other model parameters were assessed from single pili experimental data sets, it was possible to reconstruct the supposed extension response from region I and II of the last pilus even though the measurements consist of initial multi pili interactions. Figure 36 shows that the model fits pili with different lengths well, although the data consist of multi-pili interactions for short extensions. The average parameter values are presented in table 1 in Paper I.

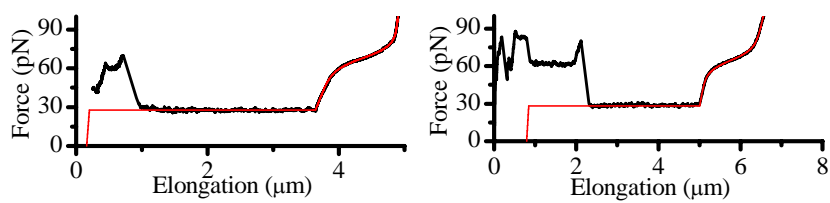

Figure 36. Two fits of the sticky-chain model to experimental data. The fit to region III gives the number of subunits where after a reconstruction of the complete pilus response can be made. As seen from the figures the fits agrees excellent with the experimental data.

### 6.2.3. Dynamic properties of P pili – Paper II

In this paper we used dynamic forces spectroscopy to probe the energy landscape of individual subunits in the PapA rod [101]. In contrast to **Paper**

**I,** in which the experiments were performed under steady-state conditions, the pilus was exposed to an increased extension velocity, $\dot{L}$, that forces the bonds to open faster than the net bond opening and closure rates. The specific measurement procedure has been described previously in chapter 6.1.3. Several DFS measurements were performed to assess the bond length, the thermal bond opening rate, and the corner velocity, i.e., the extension speed at which the rod responds with a force higher the steady-state unfolding force. The assessment of the corner velocity of a pilus gives important information of how a pilus behaves under stress and, as is discussed in **Papers VII** and **VIII**, how it possibly could be correlated to the natural environment in the urinary tract.

During DFS measurements the sequential bond opening rate is equal to the forced bond opening rate, which is written as,

$$\frac{dN_B}{dt} = \frac{\dot{L}}{\Delta x_{AB}}, \qquad (19)$$

where $\Delta x_{AB}$ is the bond opening length. Moreover, the high separation speed implies that it is possible to neglect the refolding rate, since rebinding of an open bond is very unlikely. An simplified expression that relates the unfolding force, $F_{uf}$, to the extension speed can be expressed as,

$$F_{uf} = \frac{kT}{\Delta x_{AT}} \ln\left(\frac{\dot{L}}{\Delta x_{AB} k_{AB}^{th}}\right), \qquad (20)$$

where $\Delta x_{AT}$ is the bond length and $k_{AB}^{th}$ the thermal bond opening rate. Hence, a plot of the measured mean unfolding force versus the logarithm of the extension speed gives a logarithmic dependence where the slope of the fit is given by $kT/\Delta x_{AT}$. The intersection of the line with the abscissa is a measure of the thermal extension speed, $\dot{L}^{th}$, which provides information of the thermal bond opening rate, $k_{AB}^{th} = \dot{L}^{th}/\Delta x_{AB}$. The corner velocity can be extracted from the intercept of the steady-state unfolding force and the line describing the dynamic response.

A typical dynamic response from a pilus is shown in Fig. 37. For this particular data set, a pilus was elongated at 1.1, 2.1, 4.0 8.0 15.6 30.5 69.8 117 µm/s over a distance of 3 µm, according to the procedure described in section 6.1.3. The contraction part after each event was performed under steady-state, i.e., the pilus was contracted at 0.05 µm/s to allow full refolding. The forced unfolding increased the force from 28 pN for steady-state to 34.0, 37.4, 40.4, 43.0, 46.4, 49.3 53.8 57.1 pN, for the various extension speeds. These average values are represented by the filled circles in Fig. 37, which are related to the upper x-axis which represents the speed at which the unfolding force is averaged. Moreover, these data sets were corrected with the Stokes drag force, Eq. (10), since the speed at which the pilus was elongated was high. It is clearly seen that the force increases linearly with the logarithm of the extension speed, as expect from Eq. (20).

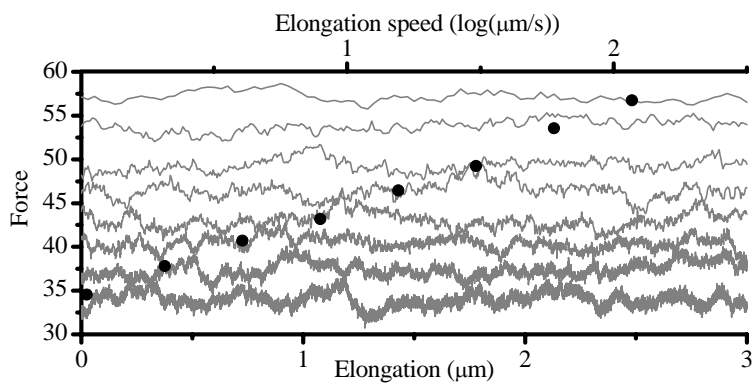

Figure 37. The data shows the force response over region II of a pilus elongated at 1.1, 2.1, 4.0 8.0 15.6 30.5 69.8 117 µm/s a distance of 3 µm, lower x-axis,. The average unfolding force at the particular speed is shown as filled circles, related to the upper x-axis. For this data set the force responses for the various speeds were average to, 34.0, 37.4, 40.4, 43.0, 46.4, 49.3 53.8 57.1 pN. It is clearly seen that the force increases linearly with the logarithm of the extension speed, as is expected from Eq. (20).

A fit with Eq. (20) to the all experimental data gave a value of 0.76 ± 0.11 nm and 0.8 ± 0.5 Hz for the bond length and the thermal bond opening rate, respectively. These values are reasonable since they are in the same magnitude as other biological bonds [64]. The balanced rate, the opening and closure of a bond, at the steady-state unfolding force was assessed to 120 Hz. Finally, the corner velocity, i.e., the speed at which the dynamic effects set in, was assessed to 0.39 ± 0.14 µm/s.

An interesting phenomenon is the difference in dynamic response between region II and III, plotted in Fig. 38. It was found, from experimental data, that region III did not respond dynamically to an increased extension speed, at least not for speeds < 100 µm/s. We believe that the reason for this is the fast response of the subunits in the pilus that act independently of their neighbors.

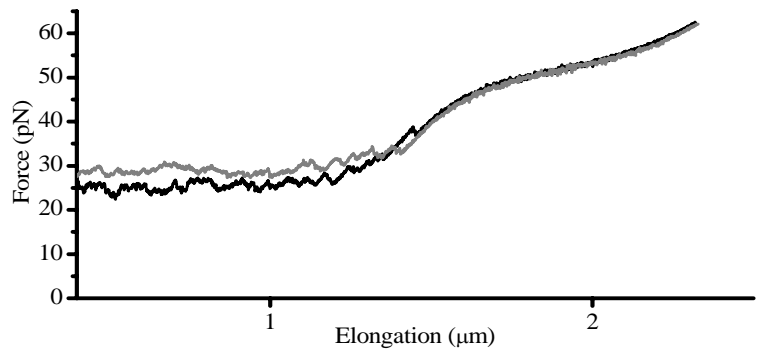

Figure 38. The data plot shows an experiment where the pilus is elongated at two different speeds. For region II, the force increases as a consequence of the forced bond opening rate, whereas the force response in region III is not affected by the increased speed. Thus, region III is still in equilibrium at this particular speed.

### 6.2.4. P pili, how tough are you? – Paper III

Several pili were exposed to repetitive extension / contraction cycles to investigate the endurance of P pili in steady-state mode. Data from such measurements is shown in Fig. 39. Panel A corresponds to the extension and Panel B to the contraction, respectively. It was found that the response was a preserved property, i.e., a P pili do not change its ability to unfold / refold, and it did not show any tendency of fatigue, even though a typical measurement lasted for > 45 min. Such measurement which lasted longer than the normal dividing rate for bacteria in natural conditions, typically ~30 min. Possibly, the resistance towards mechanical fatigue could be of importance, since pili therefore can sustain shear forces during a complete dividing cycle. To avoid bacterial division during a measurement, the samples were prepared without nutrients.

An interesting phenomenon or that is extensively discussed in a few papers is the refolding data which shows a discrete step increase in force. As seen in Fig. 39 in Panel B the high resolution of the data shows a force step in the refolding curve at the transition between region III and II. This dip is attributed to the lack of a nucleation kernel, i.e., all units are linearized and the lack of a layer prevents sequential refolding. However, when the force in the system is lowered and the bonds that stabilize the layers are getting closer, a first layer can be created. Such layer thereby provides the rest of the units to sequentially refold. An investigation of this particular phenomenon has recently been performed by Lugmaier et al. [102].

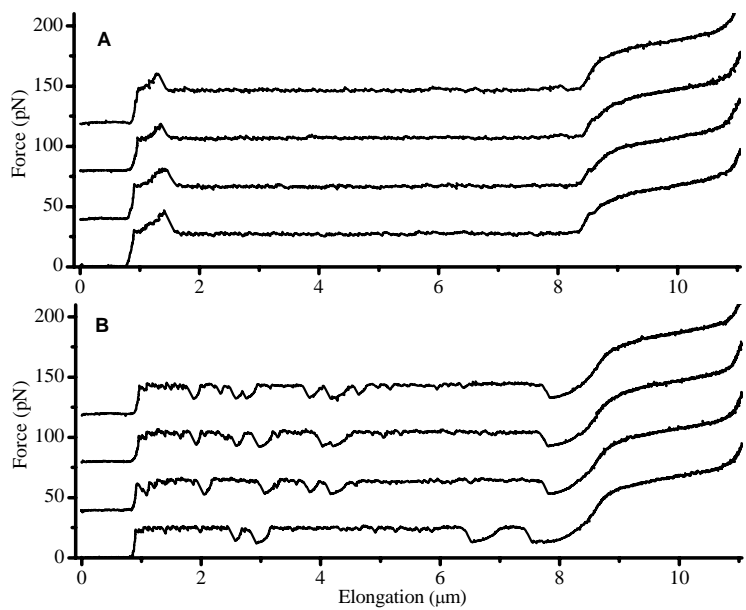

Figure 39. These measurements show how a P pilus is elongated (panel A) / contracted (panel B). The responses are from four cycles in a row, where the total time for the experiment last > 45 min. Around 40 cycles were typically performed for each set of data. A conclusion of these *in situ* repetitions are that P pili are not indicating any fatigue or conformational changes for such long and intensive measurements.

Conclusions from this work are, first, a pilus is not changing its unique quaternary extension properties after long time mechanical stress. Thus, the pilus flexibility is preserved. Secondly, a pilus must be firmly anchored to the outer cell membrane of the *E. coli* bacteria, since a pilus has never been uprooted. As a matter of fact, experiments where a single pili has been pulled with several hundreds of pN has been made [102]. Finally, a pilus will probably remain intact and show full working capabilities during and after cell division.

At last, we have studied the function of P pili after twelve hours in a sample and found that the pilus is still having the same amount of flexibility. This implies that the architecture of the pili is not depolymerised (data not shown).

### 6.2.5. Monte Carlo Simulations with a WLC description – Paper VI

Even though the sticky-chain model describes the unfolding of P pili well, a restriction is that two rate equations must be used to describe the complete force-extension response. Moreover, since bond transitions are stochastic events, because bacteria are naturally located in a liquid medium, the information regarding such events is not modeled with the analytical solution. Therefore, alternative methods, such as Monte Carlo (MC) simulations, are suitable to describe stochastic bond transitions in P pili under external stress. Moreover, a MC simulation provides a possibility to change certain conditions in the model relatively easy, e.g., to add a specific bond interaction, an adhesin interaction, or multi-pili interactions. We therefore developed a kinetic MC method to describe the unfolding / contraction response of P pili under stress.

The MC simulations were built upon a three-state energy landscape with transition probabilities between the states and a model describing the force needed to extend a flexible polymer. Depending on the external conditions, a subunit resides in one of three states, similar to the energy landscape illustrated in Fig. 30, although with an additional state. As described previously, an external force lowers the transition barriers and changes the transition probabilities. The probability of all these transitions is calculated and compared to random numbers from a uniform distribution for each sweep. For each sweep, the force is calculated in the system, the transition probabilities updated, and the state in which each bond resides in, if necessary, changed. The force in the system is calculated according to the state the monomers reside in. For units in state A, the response is linear, thus they are modeled as harmonic potentials, whereas the subunits in state B and C are modeled as connected to a semi-flexible polymer.

A common model that describes the flexibility of a semi-flexible protein polymer is the interpolated *worm like chain* (WLC) model. It has been used to describe the force-extension relationship for DNA and titin [103, 104]. The work required to stretch the polymer goes into the reduction of conformational entropy. The polymer responds thereby as an entropic spring, i.e., it shows entropic elasticity, where the force increases with extension. A well used interpolation formula for the force-extension behavior of such polymers is given by [104, 105]

$$F(x) = \frac{kT}{p}\left[\frac{1}{4}\left(\frac{1}{1-x/L_0}\right)^2 - \frac{1}{4} + \frac{x}{L_0}\right], \quad (21)$$

where $p$ is the persistence length and $L_0$ is the contour length. The persistence length is a measure of the correlation of two points along the contour of the polymer, basically how long the tangent of the two points is pointing in the same direction. Thus, the persistence length is a measure of the polymer stiffness. A polymer with short persistence length will therefore show an increase in force for short extensions since it can sample many conformations, whereas a stiff can only sample a few. A plot of Eq. (21) for three different persistence lengths, 1, 10 and 100 nm, are illustrated in Fig. 40.

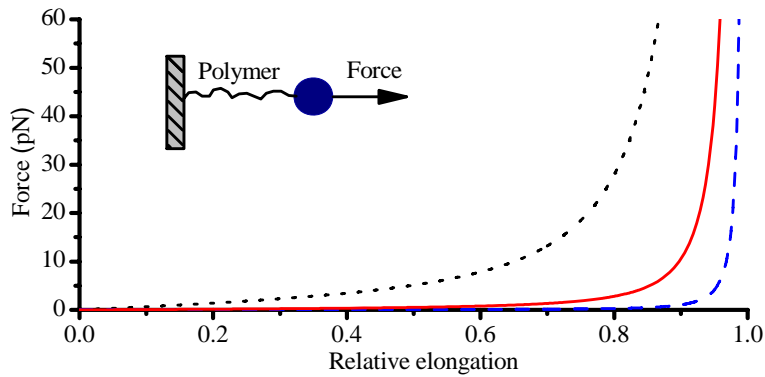

Figure 40. The three curves illustrate the force-extension relationship of the interpolated WLC. The persistence lengths are chosen to 1, 10 and 100 nm for the black (dotted), red (solid) and blue (dashed) curves, respectively.

A Monte Carlo simulation of the extension / contraction of P pili, incorporating Eq. (21) for subunits residing in state B and C, are shown in Fig. 41. As can be seen in the figures, both curves simulated under steady-state reproduce the experimental data very accurately. Notice that the figures are not from the same pilus. One can conclude from the simulations that the unfolding of the quaternary structure is a self regulating process that keeps the force constant. The load on an adhesin is therefore not constantly increasing as it would on a single bond attached to a stiff link [98]. Moreover, the accurate fit of region III, Fig. 41, shows how well the WLC model with a phase transition rate theory describes the force-extension / contraction data. The refolding of the linearized rod, i.e., reversed direction of region III, is also well represented by the simulation. The best fit of region III for P pili was obtained with a persistence length of 3.3 ± 0.6 nm. In comparison with the persistence length assessed for type 1 pili, 3.3 ± 1.6 nm [98], one can conclude that these two types of pili have comparable rigidity of their rods. This has also been supported by images [84, 85]. The deviation of the refolding of region II, i.e., the simulated data is not coinciding with the misfoldings, is only an experimental artifact and should not be consider being a error in the simulation process. Finally, the average model parameters, i.e., the values that represent the best fit to all data, are collected in table 1 in **Paper VI**.

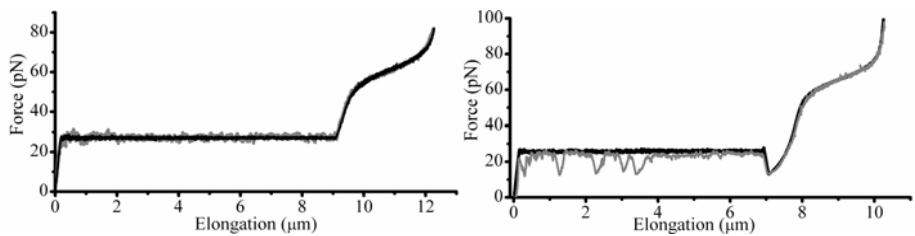

Figure 41. The figure shows how well the MC simulation agrees with the experimental data. Notice how accurately region III is model by the WLC together with a possibility to make a transition. The plots are re-plotted from [106].

## 6.3. Characterization of type 1 pili – Paper VII

Type 1 pili are one of the most studied representatives of the pilus family. They are expressed in 80 % of all wide type strains and predominantly expressed by bacteria causing cystitis, i.e., infection in the bladder. Their three dimensional structure has been investigated by scanning transmission electron microscopy and the properties of the of adhesin has been explored widely [92, 93, 107]. It has been shown that the receptor-ligand binding mediated by the adhesin is enhanced by mechanical stress. Thus, a bacteria exposed to an increased shear force will promote a stronger interaction. The mechanical properties of the rod have been measured by force measuring AFM [93, 98]. However, the low sensitivity of those measurements did not reveal all information hidden in the data. Therefore, we used FMOT to access additional information of the type 1 pili structure.

In this work, we compared the difference between P and type 1 pili under similar *in situ* conditions. The experiments showed that type 1 pili have a similar force-extension response to that of P pili as can be seen by a comparison of Fig. 33 and Fig. 42.[5] The similar response could be addressed to the similarity of the two pili, both regarding the architectures and their dimensions. However, it was a major dynamic difference between the two pili. Type 1 pili unfold the quaternary structure at ~45 pN for an extension speed of 50 nm/s (the slowest speed that our piezo stage can move), in comparison to 28 pN for P pili. Moreover, the contraction refolding force for type 1 was found to be ~27 pN, which is comparable to P pili. However, the hysteresis between the unfolding and refolding of type 1 pili indicates that the experiment was not performed at equilibrium. Thus, the assessment of the kinetics of type 1 pili needed a new approach since we were not able to assess the steady-state force due to experimental limitations. Therefore, we looked at the relaxation response, i.e., how the force decays from a dynamic response to a steady-state force. This measurement procedure is described in section 6.1.3.

---

[5] Please note that the force in Fig. 42 is decreasing far up in region III, ~4.5 µm. That is an effect of the non-linear response of the bead in the trap, originating from the fact that it is pulled too far away from the equilibrium position. However, this is not affecting the general force-extension response.

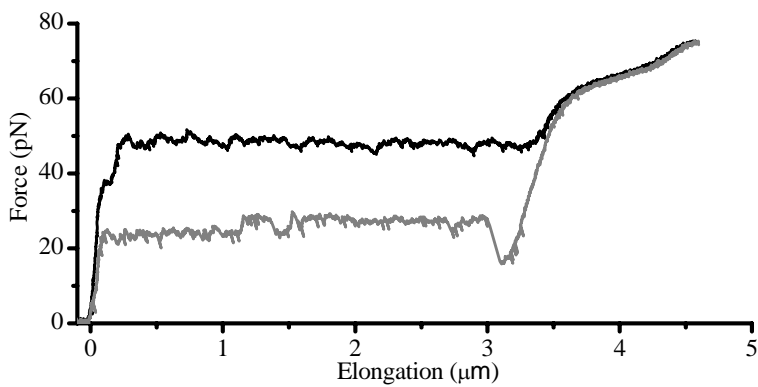

Figure 42. A response curve of type 1 pili taken at an extension speed of 0.1 µm/s. The black curve corresponds to the extension where the unfolding force is ~50 pN. The gray curve shows how the pilus refolds.

At high forces the refolding rate can be neglected and an analytical expression for how the force decays can be written as,

$$F_{\text{UF}}(t) = -\frac{kT}{\Delta x_{\text{AT}}} \ln\left[ e^{-F'\Delta x_{\text{AT}}/kT} + \frac{\Delta x_{\text{AB}} \Delta x_{\text{AT}} k_{\text{AB}}^{\text{th}}}{kT} t \right], \qquad (22)$$

where $F'$ is the initial unfolding force. A typical relaxation measurement of P and type 1 is shown in Fig. 43. The dashed curve illustrates the fit with Eq. (22) whereas the solid curve illustrates a numerical simulation in which the refolding rate has been included. Since both fits coincide for high forces it is justified to neglected the refolding rate for such fits. The force relaxes to 30 pN for type 1 pili, which is close to the steady-state force of 28 pN for P pili. However, the difference in relaxation time indicates a comparable dissimilarity in the energy landscape. The bond length in the energy landscape for type 1 pili was assessed to 0.59 nm. It can be concluded that the slow relaxation of type 1 pili is directly associated with a higher transition barrier between A and B, as compared to the response of P pili.

Another interesting observation from our data is that type 1 pili are able to refold their quaternary structure at two different force levels. As seen in Fig. 42, the force drops ~5 pN at 1.2 µm with a distinct step. We believe that this step could be correlated to the existence of an alternative subunit configuration. Thus, type 1 pili can refold with two different orientations of their subunits. This hypothesis is also supported by the fact that a force drop was most often permanent during the contraction phase. This behavior differs from that of P pili, in which such force drops have never been observed in any measurement.

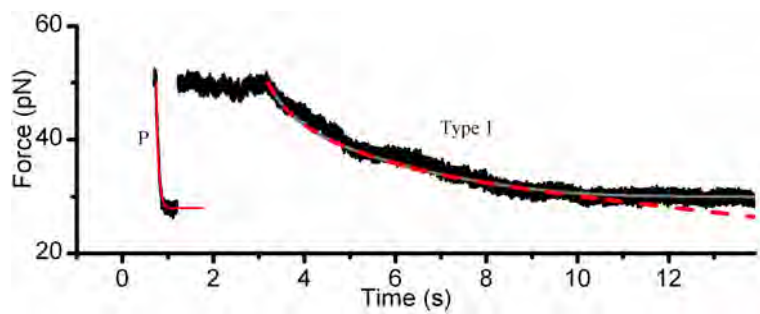

Figure 43. A typical relaxation measurement of P and type 1 pili from a dynamic response to a steady-state force. In this particular data, the P pilus and the type 1 pilus were elongated at a speed of 40 µm/s and 0.3 µm/s, respectively, which both resulted in a dynamic force of 50 pN. Thereafter the extension was stopped and the two pili decayed down to their equilibrium force, 28 and 30 pN, respectively.

In conclusion, we present a novel method to assess the kinetic parameters for a helix-like polymer that is built upon monitoring how the force relaxes down to the steady-state force. Moreover, we show that type 1 pili have a similar force-extension response to P pili, although the kinetics of the two are different. It appeared that type 1 pili have a slower transition rate, both at equilibrium and under an external force. The thermal bond opening rate was assessed to 0.0016 Hz, whereas the balanced rate, i.e., under steady-state unfolding, was assessed to 1.2 Hz. The difference in kinetics between the two pili could reflect the difference in conditions of the locations in the urinary tract at which these two are predominantly expressed. P pili, for instance, are expressed in the kidney where the shear forces are believed to be weak, whereas type 1 pili are expressed in the lower urinary tract where the shear forces are believed to be severe. Thus, our *in situ* measurements allows for a possible comparison between the kinetic function and *in vivo* relation. It could be possible that the fast dynamic force increase to shear forces optimizes the strength of the catch-bond.

Finally, the distinct force drop that was observed for type 1 pili presumably indicates that type 1 pili can refold into two alternative configurations. We have not been able to make any conclusion from that particular behavior.

## 6.4. Characterization of S pili – Paper IX

High resolution electron micrographs revealed that S pili are composite fibers protruding out from the bacterial surface [108]. The architecture of the fibers is similar to P and type 1 pili, although they are less well defined [109]. S pili are a member of the chaperone-usher pathway assembled pili family and they are predominantly associated with newborn meningitis and believed to play a role in ascending UTI [110]. Thus, characterization of their mechanical and kinetic behavior would reveal the information of the architecture and presumably it would be possible to relate its function to *in vivo* environment. Therefore, we investigated the *in situ* behavior by force measuring optical tweezers.

We measured the force-extension behavior for S pili to investigate whether the mechanical behavior was similar to P and type 1 pili. The force response of such measurement, taken under steady-state, is shown in Fig. 44. The similarity between S, P and type 1 pili response indicates that the quaternary structures are comparable. However, as mention previously, assessment of the kinetic behavior and the energy landscape parameters, i.e., how S pili behaves under dynamic conditions, reveals important information. Therefore, we measured with dynamic force spectroscopy and performed relaxation measurements at a variety of velocities.

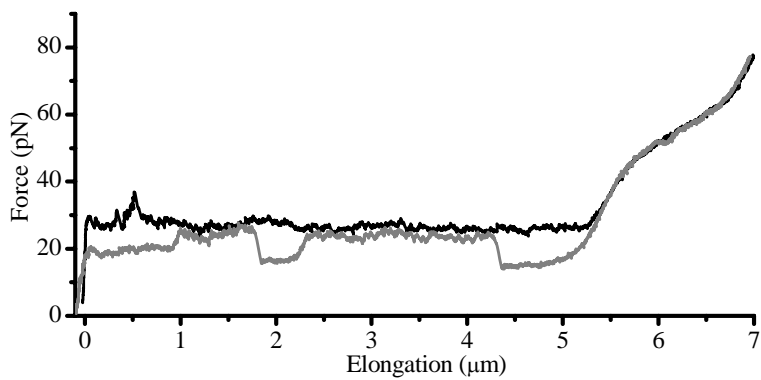

Figure 44. The force-extension curve for a single pilus measured at constant velocity. The unfolding force of the quaternary structure was 26 pN, slightly lower then P and type 1 pili.

First, a number of dynamic measurements were performed to probe the general kinetic behavior. The results show that the hysteresis between the unfolding / contraction curves is very small. This implies that the kinetics of S pili is similar to that of P pili. Therefore, the majority of our dynamic measurements were focused on the high force responses. The data suggested that we could neglect the refolding rate why it is justified to use Eq. (20) to assess the bond length and thermal extension speed. A fit to such data for four velocities, 5, 10, 20, 40 µm/s gave an average bond length of 0.66 ± 0.08 nm. Moreover, the corner velocity was assessed to 490 ± 50 nm/s whereas the thermal bond opening rate was assessed to 0.35 ± 0.15 Hz. A numerical simulation, based on these values, is shown in Fig. 45. As seen from the figure, the numerically simulated curve, where also the refolding rate is included fits the experimental data very well. In conclusion, there is a large similarity in the biomechanical behavior of S and P pili under external load. They have similar bond length and thermal bond opening rate that gives them comparable response to external force. However, our measurements indicate that S pili can refold, just as type 1 pili, at two distinct levels of force. Hence, both these pili can refold its quaternary structure in two different configurations that do not affect its unfolding properties.

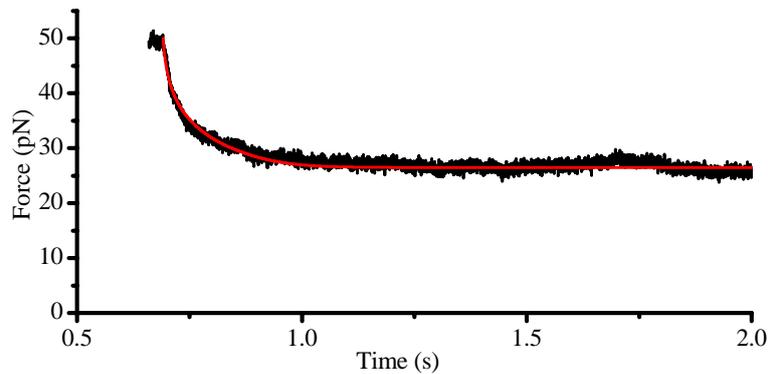

Figure 45. A relaxation fit of S pili. The response of such is similar to that of P pili, i.e., it relaxes down to a steady-state force level in ~0.2 s.

## 6.5. Review of pili properties – Paper VIII

This paper reviews the field of science on P and type 1 pili, and includes some preliminary results with P piliated bacteria in different pH and salinity buffer solutions. Moreover, we discuss the pros and cons of FMOT in comparison with AFM, i.e., the limitations and advantages of each technique in microscopic force measurements. Moreover, we show, together with chemists, how it is possible to combine pilicide treated bacteria, chemical compounds that reduce the number of expressed pili on a bacterium, and force measurements, to provide a novel biological model system that can help in assessing certain entities. For example, such model system reduces the number of pili, presumably to only a few, which is useful in specific adhesin measurements since this reduces the multi-pili interactions at the initial contact phase that otherwise affects the measurements.

Under normal environmental conditions (*in vivo*), the pH and salinity levels are fluctuating over a wide range; normally for pH 5 – 8 and for the salinity possibly between 0.038 – 1.4 mol/kg [111, 112]. It has been shown that these environmental parameters affect the gene expression [111]. However, there are to date no results of the impact of these parameters on the pili structures. Therefore, we scrutinized the influence of such for high and low pH in combination with high and low salinity levels, to mimic the *in vivo* shear forces and the extreme boundaries of pH and salinity. It was found that the unfolding properties were unaffected of the salinity. However, the refolding showed an increase in misfoldings at high salinity levels, which could possibly be correlated to screening of the layer-to-layer bonds. For the low salinity buffer, however, the refolding capability was not affected.

It was found, that the refolding was strongly affected by high pH. For pH 9.25, the number of misfoldings increased with a force-extension response similar to that with high salinity levels. At very low levels, below normal conditions, pH <3.5, there was, however, a drastic change in both the unfolding and refolding curve. At this level the entire quaternary structure seemed to be disintegrated.

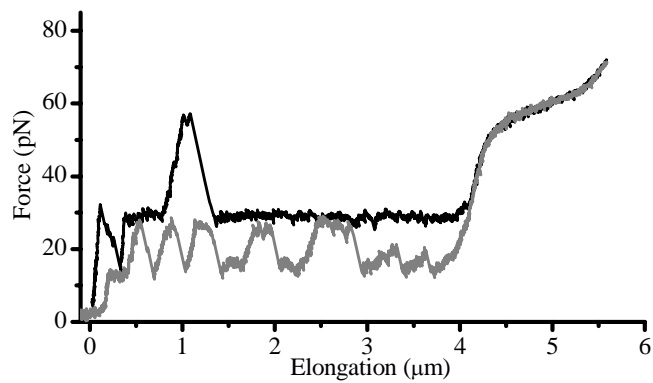

Figure 46. A pilus response under high pH. The high pH increases the number of misfoldings during the contraction. The same happens with high salinity levels. A thoroughly scrutinize of the influence of pH and salinity on the force-extension behaviour is under investigation.

# 7. ADDITIONAL STUDIES

## 7.1. Two bead optical trap – Paper V

The trapping laser in the first optical tweezers system realized in undersigned's laboratory was based upon a Ti:Sapphire laser. The laser was pumped by a 10 W Argon ion water cooled laser. The water cooling system of the laser unfortunately produced low frequency vibrations. These vibrations are picked up during force measurements as low frequency noise. Thus, long time measurements will thereby be influenced by low frequency noise. Moreover, various types of laser fluctuations; pointing instabilities, polarization-, intensity-, and mode fluctuations, can also be notable from such laser sources. Therefore, we present a dual trap technique to reduce the low frequency noise in force measuring optical tweezers. The method is built upon the concept that the noise between two simultaneously trapped beads, one used as a force transducer and one used as a reference, is correlated, since the beads are trapped and monitored with the same lasers and optical system. Therefore, the noise in the system should presumably affect both beads equally. Moreover, since the noise between the two trapped beads is correlated, it is possible to subtract the noise that comes from other sources than the object under investigation.

It was found that the low frequency noise, i.e., the mechanical and laser drifts that affect long time measurements, could be significantly reduced by this technique.

## 7.2. Integrin regulated adhesion

The first project that I got involved in, i.e., applying the force measuring optical tweezers system to a biological application, was aimed towards exploring the integrin-mediated leukocyte adhesive interactions. The adhesive leukocytes integrin receptors are important signal transduction devices relating information bidirectionally over the plasma membrane. More specifically, the dynamic regulation of cell adhesion between the LFA-1 receptor and the ICAM-1 ligand was to be investigated.

A fundamental property of the integrins is the regulation from within the cell. Although this regulation was known, there were many unanswered questions, such as; how the intracellular signals controls the extracellular activity of integrins, how the affinity of the integrins increased upon external stimuli, what are the binding strength of cell adhesion and what is the kinetics of such. Such questions could well be explored with modern force techniques, such as FMOT. However, the project was discontinued after 10 month of work due to problems of cell culture maintenance after a freezer at the Department of oncology broke down. Therefore, only preliminary results are presented in this thesis since not enough data were collected to allow for scientific conclusions with sufficient accuracy.

The experimental challenge of the project was to create a friendly environment for the leucocytes. They require a stabile temperature ~$37^O$ to survive and have a normal regulation of the integrins. Therefore a heat chamber and a heat system were developed for maintaining a natural temperature and environment for the leucocytes. A typical leukocyte cell and a typical adhesion force response are shown in Fig. 46. In comparison to the force response of pili, the response is for the leukocyte originating from a single bond.

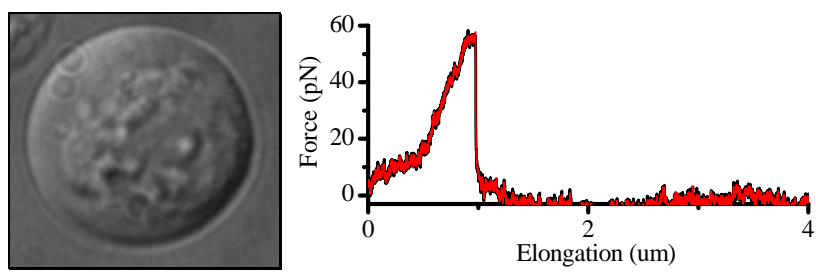

Figure 46. The left panel show a digital image of a ~10 μm leukocyte. The right panel shows a force curve from a measurement with a receptor coated bead and leukocyte. In this particular data, the maximum force is seen as a test of the bond strength.

# 8. SHORT SUMMARY AND MY CONTRIBUTION TO THE APPENDED PAPERS

## 8.1. Resume

The work performed during my PhD studies has consisted of mainly two parts, the construction of a force measuring optical tweezers instrumentation and characterization of P, S, type 1 pili. Moreover, 10 month was also spent for development of a measuring method and for the investigation of the regulation of integrins expressed by leukocytes. Although the construction of the FMOT was a time consuming project, I have not published any scientific paper purely dedicated to the instrumentation. Instead, the instrumentation has been described in the **Papers I, II, III** [3, 48, 101].

    **Paper IV** presents practical details regarding the first force measuring optical tweezers system constructed in the group and a biological model system for assessment of bacterial properties. That work was performed on an optical tweezers system that was built before I started as a PhD student.

    **Paper V** presents a method that reduces the low frequency noise produced from our water cooled Ti:Sapphire laser on the first system by a dual trap method.

    After the new instrumentation was built numerous studies of biological systems mostly related to pili have been performed, both by me and others, where **Papers I, II, III, VI, VII, VIII, IX** include those experiments that I have performed. **Papers I** and **II** introduce the sticky-chain model, a model that describes the extension and kinetic behavior of P pili. In addition, a Monte Carlo approach, built upon probability theory is presented in **Paper VI**. **Paper VII** compares P and type 1 pili and a correlation between their difference in behavior and natural environment is presented. In the quest of mapping the chaperone-usher family, S pili is investigated in **Paper IX**. Finally, **Paper VIII** reviews the field and compares our measurements on P and type 1 pili with others. Moreover, it includes the preliminary measurements of P pili in different salinity and pH buffers.

    As mentioned above, several organelles have been investigated in the field of biophysics with good results. However, an important aspect of my work is the refined measuring method that has been developed in part together with other members of the group during the last years. It allows for rapid assessment of the fundamental properties of biological interactions and the function of biological polymers. As a final point, we can nowadays, probe, investigate, and understand the fundamental properties of a new organelle in a matter of hours.

## 8.2. My contributions to the appended papers

### PAPER I – A sticky chain model of the elongation of *Escherichia coli* P pili under strain

In this work, the elongation properties of P pili are explored with force measuring optical tweezers. A stick-chain model that excellently replicates the elongation of a P pilus under steady-state conditions is presented. The model, based upon Hooks law for elongation of tandem elements and a kinetic description for bond opening and closure, is fitted to experimental data. Model parameters, such as free energy, bond opening length, and the number of monomers in a PapA rod could be assessed.

*My contribution:* Contributed to all parts of the work. I was responsible and started the research on finding a suitable model that could describe the force-extension response of P pili. The model was then discussed and refined by all of us. I constructed the optical tweezers instrumentation, made the LabView programming, all measurements and most of the analysis. The paper was written jointly by me, O. Axner and E. Fällman. The figures were made by me.

### PAPER II – Dynamic force spectroscopy of the unfolding of P pili

Dynamic force measurements reveal important information of the energy landscape of a bond. We show how the sticky-chain can be used to describe the effect of dynamic force spectroscopy on a helix-like polymer. We determined the bond length of a PapA subunit, the thermal elongation speed, and the corner velocity, i.e., the elongation speed at which the rod responds with an increased force. Thus, we probed the energy landscape and determined the important entities of such. Finally, we explain and show why a pilus responds different to stress in region II and III.

*My contribution:* Contributed to all parts of the work. Again, the measurements were performed on the system I have constructed including the LabView programming. I performed all measurements and most of the analysis. I made all figures except one and wrote the paper together with Ove Axner and Erik Fällman.

PAPER III – **Force measuring optical tweezers system for long time measurements of Pili stability**

This paper presents the properties of the optical tweezers system constructed by undersigned together with long time measurements of P pili. The durability of P pili was scrutinized with repetitive elongation / contraction measurements that were pursued for more than > 45 min. The work shows, in contrast to [78], that the pili structure is extremely durable and that it can be unfolded and refolded multiple times without losing its intrinsic properties. Moreover, the results indicate that a pilus must be firmly anchored to the outer cell membrane of *E. coli* bacteria, since the pilus was not uprooted during such long measurements. In addition, such long measurements indicate that a pilus will remain intact and show full working capabilities probably during cell and after cell division.

*My contribution:* Contributed to all parts of the work. Again, the measurements were performed on the system I have constructed including the LabView programming. I performed all measurements and most of the analysis. Most of the paper was written by undersigned and Ove Axner, whereas all figures were made by undersigned.

PAPER IV – **Optical tweezers based force measurement system for quantitating binding interactions: system design and application for the study of bacterial adhesion**

This invited paper describes the system design of the "old" optical tweezers instrumentation and a detailed biological assay for measurements of bacterial pili.

*My contribution:* Since this system was constructed before I started as a PhD student, I only participated in completing the manuscript. I made some of the figures.

PAPER V – **A dual trap technique for reduction of low frequency noise in force measuring optical tweezers**

Long time measurements are often influenced by mechanical drifts and laser fluctuations. Therefore, we present a dual trap technique to reduce the low frequency noise in force measuring optical tweezers. The method uses two trapped beads, one used as a force transducer and one used as a reference. The beads were trapped and monitored with the same lasers and optical system. Therefore, the noise in the system will affect both beads equally. However, by correlating the noise between the two trapped beads, it was possible to subtract the noise that comes from other sources than the object under investigation. It was found that the low frequency noise was significantly reduced for low frequencies, i.e., the mechanical and laser drifts that affect long time measurements.

*My contribution:* The experimental setup of this project was mostly performed by author 1 as a master project. However, I assisted him with parts of the experimental setup. Moreover, I contributed to parts of the writing and made some figures.

PAPER VI – **Modelling of the Elongation and Retraction of *E. Coli* P pili under Strain by Monte Carlo simulations**

Analytical models for opening and closing, e.g., the sticky-chain model, do not include stochastic effects. Therefore, an alternative method based upon Monte Carlo simulations with a worm like chain description of the linearized rod was developed. We determined the persistence length of the pili by fitting Monte Carlo simulations to experimental data. The simulations were found to be consistent with the steady-state sticky-chain model and the dynamic effect of region II. We also investigated the dynamic effect of region III with the simulation.

*My contribution:* Contributed to all parts of the work. Again, the measurements were performed on the system I have constructed including the LabView programming. I performed all measurements. I also contributed to parts of the theoretical work and the paper was written jointly by all authors. I was the contact author of the paper responsible for the manuscript and submission.

### PAPER VII – The biomechanical properties of *E. coli* pili for urinary tract attachment reflect the host environment

Bacteria are carriers of multiple pili genes and the expression of these are steered by the environment. This paper compares the differences of P and type 1 pili, which are expressed in pyelonephritis and cystitis, respectively. It is shown that type 1 pili have a similar force-extension response to P pili. However, the kinetics of type 1 pili shows a significant different dynamic response to increased extension speeds. We therefore present a novel technique to assess the bond length and thermal bond opening rate from the relaxation data. The difference in kinetics between the two pili could reflect the location in the urinary tract at which these two are expressed. Moreover, the measurements revealed that type 1 pili can refold at two different forces. We believe that this force difference reflects an alternative refolding configuration.

*My contribution:* Contributed to all parts of the work. Again, the measurements were performed on the system I have constructed including the LabView programming. I performed all measurements and most of the analysis. I made a majority of the figures and wrote the paper together with Erik Fällman.

### PAPER IX – Physical properties of biopolymers assessed by optical tweezers

This paper reviews the field and highlights the use of optical tweezers for accurate force measurements. It shows preliminary measurements at low and high pH and salinity buffer solutions. It was found that the refolding was strongly affected by both low and high pH, with an increased number of misfoldings. In addition, at very low pH there was a drastic change in both the unfolding and refolding curve. At this level the entire quaternary structure seemed to be denaturized.

*My contribution:* Contributed to all parts of the work. Again, the measurements were performed on the system I have constructed including the LabView programming. I performed some of the measurements and parts of the analysis. The paper was written by all authors.

## PAPER IX – **Characterization of S pili**

In the quest of characterizing the pili family, S pili is investigated and analyzed with the FMOT. It was found that S pili have a similar force response to both other types of pili studied. S and P pili have similar dynamic response, which reflects the similar bond and thermal rates. In addition, S and type 1 pili are both able to refold at two levels of force. Thus, they can refold at an alternative configuration, which has never been observed for P pili. Finally, it has been shown that the adhesin of type 1 pili is of catch bond type, i.e., bonds that are enhanced upon force. However, the general consensus is that P pili are slip bonds. Thus, the similarity between S and P pili could imply that S also is of slip bond type.

*My contribution:* Contributed to all parts of the work. Again, the measurements were performed on the system I have constructed including the LabView programming. The authors contributed evenly to the measurements and writing of the manuscript.

# 9. ACKNOWLEDGEMENTS

**Ove Axner** - for being an excellent supervisor, always showing a positive mind, providing brilliant ideas, a great teacher both during the years as undergraduate and graduate student, and finally showing a huge sense of humour that lightens the days.

**Erik Fällman** – for teaching me the exciting field of Optical tweezers, helping me in the lab, and for being a really good friend. I am very grateful that I had the opportunity to work with you.

**TFE guys, Oscar Björnham** (Mr. Matlab) – for all his help with programming and for being a great companion and **Staffan Schedin** – for explaining difficult things in a very instructive manner.

**Bernt-Eric Uhlin** and **Fredrik Almqvist** – for the interdisciplinary collaboration and all help in the field of microbiology / chemistry.

**Jörgen Eriksson** – for always helping without any hesitation, being a great friend and for all quality training time.

A super thanks to the people in the workshop who help me with electronics, construction and modification of various odd things; **Tomas Gustafsson**, **Lena Åström**, **Lars Karlsson** and **Martin Forsgren**.

**Ann-Charlott**, **Katarina**, **Lilian,** and **Margaretha** for all help with administration and other bureaucratic stuff.

I have had the pleasure to teach together with all of the members of the laser cooling group. Therefore, a great thanks for these years of fun; **Peder Sjölund**, **Rober Saers**, **Magnus Rehn**, **Henning Hagman**, **Martin Zélan** and the group leader **Anders Kastberg**. By the way, how do you make your Mac go faster? Drop it from a higher window! Lucky you, your on the second floor, hehe.

**Krister Wiklund** and **Martin Servin** – Thanks for the fruitful collaboration so far and for being two great guys from the third floor.

**Andrzej Dzwilewski** – for being a great roommate and for teaching me polish.

**The Physics staff –** Florian Schmidt, Alexandra Foltynowicz, Patrik Stenmark, Piotr Matyba, Sebastian Bernhardsson, Mats Forsberg, Joakim Lundin, Annie Reiniusson, Jens Zamanian, Mats Svantesson Maria Tranbeck, Andreas Grönlund, Michael Bradley, Lars-Erik Svensson, Leif Hassmyr, Maria Hamrin, Mattias Marklund, Patrik Norqvist, Alexandr Talyzin, Ludvig Edman, Thomas Wågberg, Mats Nyhlên, Hans Forsman, Ove Andersson, Bertil Sundqvist, Sylvia Benckert, Sune Pettersson and all others at the department of physics.

My lovely girlfriend **Ann-Sofie** and my gorgeous daughter **Nea**.

All my **relatives** and **friends**.

**OHC** – The Crew! Thanks for all support guys and I am already preparing for the next party! OHC, står när de andra faller.

82. Johnson, J.R. and T.A. Russo, *Uropathogenic Escherichia coli as agents of diverse non-urinary tract extraintestinal infections.* Journal of Infectious Diseases, 2002. **186**(6): p. 859-864.
83. Russo, T.A. and J.R. Johnson, *Medical and economic impact of extraintestinal infections due to Escherichia coli: focus on an increasingly important endemic problem.* Microbes and Infection, 2003. **5**(5): p. 449-456.
84. Gong, M.F. and L. Makowski, *Helical structure of P pili from Escherichia coli - Evidence from X-ray fiber diffraction and scanning-transmission electron-microscopy.* Journal of Molecular Biology, 1992. **228**(3): p. 735-742.
85. Saulino, E.T., E. Bullitt, and S.J. Hultgren, *Snapshots of usher-mediated protein secretion and ordered pilus assembly.* Proceedings of the National Academy of Sciences of the United States of America, 2000. **97**(16): p. 9240-+.
86. Bullitt, E. and L. Makowski, *Bacterial adhesion pili are heterologous assemblies of similar subunits.* Biophysical Journal, 1998. **74**(1): p. 623-632.
87. Bullitt, E., et al., *Development of pilus organelle subassemblies in vitro depends on chaperone uncapping of a beta zipper.* Proceedings of the National Academy of Sciences of the United States of America, 1996. **93**(23): p. 12890-12895.
88. Lindberg, F., et al., *Localization of the Receptor-Binding Protein Adhesin at the Tip of the Bacterial Pilus.* Nature, 1987. **328**(6125): p. 84-87.
89. Lindberg, F., B. Lund, and S. Normark, *Gene-Products Specifying Adhesion of Uropathogenic Escherichia-Coli Are Minor Components of Pili.* Proceedings of the National Academy of Sciences of the United States of America, 1986. **83**(6): p. 1891-1895.
90. Dodson, K.W., et al., *Structural basis of the interaction of the pyelonephritic E. coli adhesin to its human kidney receptor.* Cell, 2001. **105**(6): p. 733-743.
91. Hung, C.S., et al., *Structural basis of tropism of Escherichia coli to the bladder during urinary tract infection.* Molecular Microbiology, 2002. **44**(4): p. 903-915.
92. Thomas, W.E., et al., *Bacterial adhesion to target cells enhanced by shear force.* Cell, 2002. **109**(7): p. 913-923.
93. Forero, M., et al., *Uncoiling Mechanics of Escherichia coli Type I Fimbriae Are Optimized for Catch Bonds.* PLoS Biology, 2006. **4**(9): p. 1509-1516.
94. Andersson, M., et al., *Physical Properties of Biopolymers Assessed by Optical Tweezers: Analyses of Bacterial Pili.* Chemphyschem, 2007: p. In press.
76

# 11. Unsolved questions

**Graduate level**

|   | 3 | 2 |   |   |   | 9 | 4 |   |
|---|---|---|---|---|---|---|---|---|
| 5 |   |   |   |   |   | 7 |   |   |
|   |   |   | 8 |   |   | 3 |   | 1 |
|   |   |   |   |   | 5 | 8 |   |   |
|   |   | 1 | 6 |   | 4 |   |   |   |
| 9 |   | 7 | 1 |   |   |   |   |   |
|   | 2 |   |   |   | 1 |   |   | 4 |
|   |   |   |   | 4 | 8 | 5 |   |   |
| 3 |   |   | 5 |   |   |   |   |   |

**Professor level**

| 7 | 6 |   |   | 4 | 5 | 1 | 9 | 8 |
|---|---|---|---|---|---|---|---|---|
|   |   | 1 |   | 8 |   |   |   |   |
|   |   |   |   |   | 1 | 5 | 2 |   |
|   |   | 9 | 4 |   | 2 | 8 | 6 |   |
|   |   |   | 5 |   | 3 | 9 |   |   |
|   | 7 |   | 8 |   |   | 4 |   |   |
|   |   |   | 1 | 2 | 8 | 3 |   |   |
| 2 | 1 |   |   | 5 | 4 |   | 8 | 9 |
|   | 4 | 8 | 6 |   |   | 2 |   |   |